\documentclass{article}

\usepackage{microtype}
\usepackage{graphicx}
\usepackage{subcaption}
\usepackage{booktabs}
\usepackage{hyperref}

\usepackage[accepted]{icml2026}
\makeatletter
\renewcommand{\ICML@appearing}{\textit{Accepted at Learning to Listen: ICML 2026
Workshop on Machine Learning for Audio} (non-archival). Copyright 2026 by the
author(s).}
\makeatother

\usepackage{amsmath}
\usepackage{amssymb}
\usepackage{mathtools}
\usepackage{amsthm}
\usepackage{multirow}
\usepackage{tikz}
\usetikzlibrary{positioning, arrows.meta, fit, calc, backgrounds,
decorations.pathreplacing}

\usepackage[capitalize,noabbrev,nameinlink]{cleveref}

\usepackage{amsmath,amsfonts,bm}

\def\eqref#1{equation~\ref{#1}}

\def\1{\bm{1}}

\def\vzero{{\bm{0}}}

\def\va{{\bm{a}}}

\def\vv{{\bm{v}}}

\def\vx{{\bm{x}}}

\def\vz{{\bm{z}}}

\def\mI{{\bm{I}}}

\DeclareMathAlphabet{\mathsfit}{\encodingdefault}{\sfdefault}{m}{sl}
\SetMathAlphabet{\mathsfit}{bold}{\encodingdefault}{\sfdefault}{bx}{n}

\graphicspath{{./}{./figures/}{./attention_diagrams/}}

\usepackage{xspace}
\newcommand{\method}{AV-JEPA\xspace}

\renewcommand{\paragraph}[1]{\noindent\textbf{#1}~~}

\crefname{section}{Sec.}{Secs.}
\crefname{appendix}{App.}{Apps.}
\crefname{algorithm}{Alg.}{Algs.}
\crefname{figure}{Fig.}{Figs.}

\icmltitlerunning{AV-JEPA: Extending LeJEPA to Audio-Visual Self-Supervised
Learning}

\begin{document}

\twocolumn[
  \icmltitle{AV-JEPA: Extending LeJEPA to\\
    Audio-Visual Self-Supervised Learning}

  \icmlsetsymbol{equal}{*}

  \begin{icmlauthorlist}
    \icmlauthor{Benjamin Robson}{aalto}
    \icmlauthor{Santeri Mentu}{aalto}
    \icmlauthor{Wenshuai Zhao}{aalto}
    \icmlauthor{Arno Solin}{aalto}
  \end{icmlauthorlist}

  \icmlaffiliation{aalto}{ELLIS Institute Finland and Department of Computer Science, Aalto University, Espoo, Finland}

  \icmlcorrespondingauthor{Benjamin Robson}{benjamin.robson@aalto.fi}

  \icmlkeywords{Self-supervised learning, audio, video, JEPA, multimodal}

  \vskip 0.3in
]

\printAffiliationsAndNotice{}

\begin{abstract}

We present \method, an elegant multimodal extension of LeJEPA to audio-visual
self-supervised learning. Using an early-fusion Vision Transformer and
\emph{modality dropout as masking}, the model is trained to align the embeddings
of global and per-modality local views, while the SIGReg objective encourages a
theoretically optimal distribution. This achieves cross-modal alignment in the
latent space, resulting in a remarkably clean architecture with no decoder, EMA
teacher, complex multi-term losses, or contrastive negatives. The proposed
AV-JEPA backbone delivers competitive classification performance on VGGSound
(57.1\% top-1) and AudioSet (32.7 mAP) and supports zero-shot audio-video
retrieval out of the box.
\looseness-1

\end{abstract}

\section{Introduction}
\label{sec:intro}

Self-supervised learning (SSL) has become the dominant paradigm for learning
high-quality representations from unlabelled data. In the audio-visual domain,
the prevailing approach is \emph{masked autoencoding}: models such as
AV-MAE~\cite{georgescu2023avmae}, CAV-MAE~\cite{gong2023cavmae},
MAViL~\cite{huang2023mavil}, and CAV-MAE Sync~\cite{araujo2025cavmaesync}
reconstruct masked audio spectrograms and video patches through dedicated
decoder networks. While effective, these methods lack formal guarantees on
embedding quality and require architectural overhead such as separate decoders,
carefully tuned masking ratios, or contrastive negative-pair losses.

An alternative paradigm, the \emph{Joint-Embedding Predictive Architecture}
\citep[JEPA,][]{lecun2022path}, avoids reconstruction entirely by operating in
the latent space: the model learns to predict the embedding of one view from
another. Recently, LeJEPA~\cite{balestriero2025lejepa} established a rigorous
theoretical foundation for JEPAs by proving that the isotropic Gaussian is the
uniquely optimal embedding distribution for minimizing downstream prediction
risk, and enforces it via \emph{Sketched Isotropic Gaussian Regularization}
(SIGReg). However, LeJEPA has only been validated on single-modality vision
tasks.

Concretely, we ask: \emph{`Can a JEPA achieve cross-modal audio-visual
alignment, and in particular yield strong audio representations, without a
decoder, contrastive negatives, EMA teachers, stop-gradients, or per-modality
pretraining?'} We extend LeJEPA to the audio-visual setting with
\textbf{\method}. Our key contributions are: {\em (i)}~an extension of
JEPA-based self-supervised learning to audio-visual representation learning;
{\em (ii)}~\emph{cross-modal view generation}, where local views alternate
between audio-only and video-only inputs (the other modality zeroed), creating
an implicit cross-modal prediction task entirely in latent space; {\em (iii)}~an
early-fusion ViT architecture that processes both modalities jointly through a
single shared transformer; and {\em (iv)}~empirical results from our main
experiment, AudioSet-2M pretraining followed by fine-tuning, showing that the
resulting backbone reaches \textbf{57.1\%} top-1 on VGGSound and \textbf{32.7
mAP} on AudioSet, with a single-modality breakdown confirming a strongly
audio-driven representation, and supports cross-modal retrieval out of the box.
Qualitatively, meaningful cross-modal attention to the sound source emerges
purely from the JEPA objective, without any localization supervision.

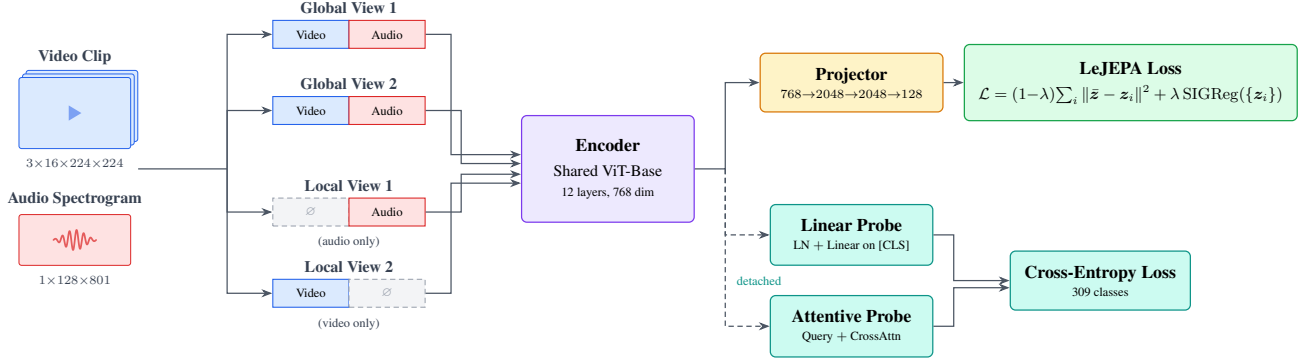
\begin{figure*}[t]
  \centering
%

\definecolor{vidfill}{HTML}{DBEAFE}
\definecolor{vidbord}{HTML}{2563EB}
\definecolor{audfill}{HTML}{FEE2E2}
\definecolor{audbord}{HTML}{DC2626}
\definecolor{absfill}{HTML}{F3F4F6}
\definecolor{absbord}{HTML}{9CA3AF}
\definecolor{encfill}{HTML}{EDE9FE}
\definecolor{encbord}{HTML}{7C3AED}
\definecolor{projfill}{HTML}{FEF3C7}
\definecolor{projbord}{HTML}{D97706}
\definecolor{lossfill}{HTML}{DCFCE7}
\definecolor{lossbord}{HTML}{16A34A}
\definecolor{probefill}{HTML}{CCFBF1}
\definecolor{probebord}{HTML}{0D9488}
\definecolor{inputfill}{HTML}{F9FAFB}
\definecolor{inputbord}{HTML}{D1D5DB}
\definecolor{gtext}{HTML}{374151}
\definecolor{garrow}{HTML}{4B5563}

\resizebox{\textwidth}{!}{%
\begin{tikzpicture}[
    >=Stealth,
    arr/.style={->, line width=0.7pt, garrow},
    arrdash/.style={->, line width=0.7pt, garrow, densely dashed},
    block/.style={
        rectangle, rounded corners=3pt, line width=0.7pt,
        font=\normalsize, align=center,
        minimum height=1.2cm, inner sep=8pt,
    },
    modbar/.style={
        rectangle, minimum width=1.5cm, minimum height=0.55cm,
        line width=0.6pt, font=\scriptsize, align=center,
    },
    viewlabel/.style={font=\footnotesize\bfseries, gtext},
    dim/.style={font=\scriptsize\color{gtext}, align=center},
]

\foreach \off in {0.14, 0.07, 0} {
    \draw[vidbord, fill=vidfill, line width=0.5pt, rounded corners=1.5pt]
        (-1.1+\off, 1.5+\off) rectangle (1.1+\off, 2.9+\off);
}
\fill[vidbord!60] (-0.12, 2.05) -- (-0.12, 2.35) -- (0.13, 2.2) -- cycle;

\begin{scope}
\clip[rounded corners=1.5pt] (-1.1, -0.80) rectangle (1.1, 0.20);
\fill[audfill] (-1.1,-0.80) rectangle (1.1,0.20);

  \begin{scope}[xshift=0pt, yshift=-0.30cm, xscale=0.42, yscale=0.20]
  \draw[audbord!75, line width=1.0pt, line cap=round, line join=round]
    plot[domain=-1.0:1.0, samples=160, smooth]
      (\x, {sin(\x*1080) * (0.55 + 0.45*cos(\x*180))});
  \end{scope}
\end{scope}
\draw[audbord, line width=0.5pt, rounded corners=1.5pt]
    (-1.1, -0.80) rectangle (1.1, 0.20);

\node[font=\footnotesize\bfseries, gtext] at (0, 3.3)
    {Video Clip};
\node[font=\footnotesize\bfseries, gtext] at (0, 0.50)
    {Audio Spectrogram};
\node[dim] at (0, 1.20) {$3{\times}16{\times}224{\times}224$};
\node[dim] at (0, -1.10) {$1{\times}128{\times}801$};

\node[inner sep=0pt, minimum size=0pt] (input) at (1.24, 1.1) {};

\def\vx{5.4}

\node[viewlabel] at (\vx, 4.28) {Global View 1};
\node[modbar, draw=vidbord, fill=vidfill] (g1v) at (\vx-0.75, 3.75) {Video};
\node[modbar, draw=audbord, fill=audfill] (g1a) at (\vx+0.75, 3.75) {Audio};

\node[viewlabel] at (\vx, 2.78) {Global View 2};
\node[modbar, draw=vidbord, fill=vidfill] (g2v) at (\vx-0.75, 2.25) {Video};
\node[modbar, draw=audbord, fill=audfill] (g2a) at (\vx+0.75, 2.25) {Audio};

\node[viewlabel] at (\vx, 0.78) {Local View 1};
\node[modbar, draw=absbord, fill=absfill, densely dashed]
    (l1v) at (\vx-0.75, 0.25) {\color{absbord}$\varnothing$};
\node[modbar, draw=audbord, fill=audfill]
    (l1a) at (\vx+0.75, 0.25) {Audio};
\node[dim] at (\vx, -0.35) {(audio only)};

\node[viewlabel] at (\vx, -0.82) {Local View 2};
\node[modbar, draw=vidbord, fill=vidfill]
    (l2v) at (\vx-0.75, -1.35) {Video};
\node[modbar, draw=absbord, fill=absfill, densely dashed]
    (l2a) at (\vx+0.75, -1.35) {\color{absbord}$\varnothing$};
\node[dim] at (\vx, -1.95) {(video only)};

\coordinate (splitpt) at (3.0, 1.1);
\draw[-, line width=0.7pt, garrow] (input.east) -- (splitpt);
\draw[arr] (splitpt) |- (g1v.west);
\draw[arr] (splitpt) |- (g2v.west);
\draw[arr] (splitpt) |- (l1v.west);
\draw[arr] (splitpt) |- (l2v.west);

\node[block, draw=encbord, fill=encfill,
      minimum width=3.4cm, minimum height=2.1cm]
    (enc) at (10.5, 1.1)
    {\textbf{Encoder}\\[2pt]
     {\small Shared ViT-Base}\\[-1pt]
     {\scriptsize 12 layers, 768 dim}};

\draw[arr] (g1a.east) -- ++(0.5,0) |- ([yshift=8pt]enc.west);
\draw[arr] (g2a.east) -- ++(0.7,0) |- ([yshift=3pt]enc.west);
\draw[arr] (l1a.east) -- ++(0.7,0) |- ([yshift=-3pt]enc.west);
\draw[arr] (l2a.east) -- ++(0.5,0) |- ([yshift=-8pt]enc.west);

\node[block, draw=projbord, fill=projfill, minimum width=3.6cm, minimum height=0.95cm]
    (proj) at (15.3, 2.8)
    {\textbf{Projector}\\[-2pt]{\scriptsize $768 {\to} 2048 {\to} 2048 {\to} 128$}};

\node[block, draw=probebord, fill=probefill, minimum width=3.2cm, minimum height=0.95cm]
    (linprobe) at (15.3, -0.2)
    {\textbf{Linear Probe}\\[-2pt]{\scriptsize LN $+$ Linear on [CLS]}};

\node[block, draw=probebord, fill=probefill, minimum width=3.2cm, minimum height=0.95cm]
    (attprobe) at (15.3, -2.0)
    {\textbf{Attentive Probe}\\[-2pt]{\scriptsize Query $+$ CrossAttn}};

\draw[arr] (enc.east) -- ++(0.6,0) coordinate (esplit)
    |- (proj.west);

\draw[arrdash] (esplit) |- (linprobe.west);
\draw[arrdash] (esplit) |- (attprobe.west);

\node[font=\scriptsize\color{probebord}, anchor=west]
    at ([xshift=3pt]esplit |- {$(linprobe.west)!0.5!(attprobe.west)$})
    {detached};

\node[block, draw=lossbord, fill=lossfill,
      minimum width=4.0cm, minimum height=1.5cm]
    (loss) at (20.8, 2.8)
    {\textbf{LeJEPA Loss}\\[2pt]
     {\footnotesize
      $\mathcal{L} = (1{-}\lambda)\!\sum_{i}\|\bar{\vz} - \vz_i\|^2
       + \lambda\,\mathrm{SIGReg}(\{\vz_i\})$}};

\node[block, draw=probebord, fill=probefill,
      minimum width=2.8cm, minimum height=0.95cm]
    (celoss) at (20.2, -1.1)
    {\textbf{Cross-Entropy Loss}\\[-2pt]{\scriptsize 309 classes}};

\draw[arr] (proj) -- (loss);
\draw[arr] (linprobe.east) -- ++(0.4,0) coordinate (pturn) |- (celoss.west);
\draw[arr] (attprobe.east) -- (pturn |- attprobe.east) |- ([yshift=-4pt]celoss.west);

\end{tikzpicture}}
  \caption{\textbf{\method training pipeline.} Each clip is split into $G{=}2$
  global views (both modalities) and $K{=}2$ local views (alternating audio-only
  / video-only, the absent modality zeroed). All views go through a shared
  ViT-Base early-fusion encoder over video tubelets and audio mel-spectrogram
  patches. The LeJEPA loss pulls every view embedding toward the joint-modality
  center~$\bar{\vz}$ while SIGReg enforces an isotropic Gaussian embedding
  distribution. During VGGSound pretraining we additionally attach detached
  linear and attentive classification probes.}
  \label{fig:pipeline}
\end{figure*}

\section{Methods}
\label{sec:method}
\method adapts LeJEPA to cross-modal audio-visual learning through {\em (i)}~an
early-fusion architecture that embeds both modalities into a single token
sequence, and {\em (ii)}~a view-generation strategy that uses modality dropout
as the partial-view mechanism. The full pipeline is shown in
\cref{fig:pipeline}.

\paragraph{LeJEPA loss}
LeJEPA~\cite{balestriero2025lejepa} shows that the isotropic Gaussian
$\mathcal{N}(\vzero,\mI)$ is the optimal embedding distribution for both linear
and nonlinear downstream probes, and enforces it through SIGReg: a sliced
characteristic-function test that projects embeddings onto $M$ random unit-norm
directions and matches each univariate projection to the Gaussian target via
Epps--Pulley.  Given $G$ global and $K$ local views with embeddings $\vz_i =
\mathrm{Proj}(f_\theta(\vx_i))$ and the joint-view center
$\bar{\vz}=\tfrac{1}{G}\sum_{g=1}^{G}\vz_g$, the LeJEPA loss is
\begin{equation}
\label{eq:lejepa}
\mathcal{L} = (1-\lambda)\,\underbrace{\frac{1}{G+K}\sum_{i=1}^{G+K}\|\bar{\vz}-\vz_i\|^2}_{\text{invariance}}
\;+\;\lambda\,\underbrace{\vphantom{\sum_{i=1}^{G+K}}\mathrm{SIGReg}(\{\vz_i\})}_{\text{regularization}},
\end{equation}
with a single trade-off scalar $\lambda$.

\paragraph{Audio-video early-fusion encoder}
Raw audio is resampled to 16\,kHz and converted to a $1{\times}128{\times}801$
mel spectrogram (128 mel bins, 801 time frames from an 8\,s clip); a
$16{\times}16$ Conv2D patch embedding yields $8{\times}50{=}400$ audio tokens
with factorized (frequency, time) positional embeddings. Video frames of shape
$3{\times}T{\times}224{\times}224$ ($T{=}16$) are tokenized by a
$2{\times}16{\times}16$ Conv3D tubelet, giving $1568$ video tokens with
factorized (spatial, temporal) positional embeddings. A learnable \texttt{[CLS]}
token is prepended:
\begin{equation}
[\texttt{CLS};\,\vv_1,\ldots,\vv_{1568};\,\va_1,\ldots,\va_{400}],
\end{equation}
with learned modality-type embeddings (ID 0 = video, ID 1 = audio) added to
distinguish modalities. The full 1969-token sequence is processed by a
ViT-Base~\cite{dosovitskiy2021vit} encoder (12L, $d{=}768$, 12 heads). The
\texttt{[CLS]} output is projected by a 3-layer MLP
($768{\to}2048{\to}2048{\to}128$, BatchNorm, GELU) before the LeJEPA loss is
applied. See \cref{fig:encoder} for the full encoder schematic.

\paragraph{Modality dropout as partial-view mechanism}
Standard LeJEPA generates view diversity through spatial augmentations. We add
an audio-visual axis: each 10\,s clip is split into two temporally offset 8\,s
crops, used to form two global views (both modalities present, light
augmentation) and two local views, one \textbf{audio-only} (video tokens zeroed)
and one \textbf{video-only} (audio tokens zeroed, video with standard
augmentations). The invariance term in \cref{eq:lejepa} then pushes each
single-modality embedding toward the joint-modality center $\bar{\vz}$, so the
model must learn, from audio alone, an embedding predictive of the joint
audio-video representation (and vice versa). Cross-modal alignment happens
entirely in latent space, with no decoder or reconstruction target. SIGReg
simultaneously prevents collapse of these dropout-induced embeddings.

\paragraph{Online probing}
On the labelled VGGSound dataset, we additionally train two classification heads
on \emph{detached} backbone features: a linear probe (LayerNorm + linear on
\texttt{[CLS]}) and an attentive probe (one learnable query, 12-head
cross-attention over patch tokens), both with cross-entropy and label smoothing
0.1 at learning rate $10^{-3}$.

\section{Experiments}
\label{sec:experiments}

Our \textbf{main experiment} pretrains \method on
AudioSet-2M~\cite{gemmeke2017audioset} and fine-tunes the resulting backbone on
VGGSound~\cite{chen2020vggsound} for audio-visual classification
(\cref{sec:exp_vggsound}). We additionally fine-tune the same
AudioSet-pretrained backbone on AudioSet itself (\cref{sec:exp_audioset}) as a
sanity check that pretraining transfers to its source distribution, and probe
cross-modal retrieval (\cref{sec:retrieval}) as a complementary check that the
learned embedding space is genuinely shared across modalities. As a controlled
secondary study, we also pretrain (and fine-tune) on VGGSound alone, isolating
the contribution of the JEPA objective from data scale. All runs share the same
ViT-Base early-fusion encoder and LeJEPA recipe, differing only in pretraining
dataset, batch size, and training budget.

\paragraph{Datasets}
\textbf{AudioSet}~\cite{gemmeke2017audioset}: ${\sim}$2M YouTube clips,
multi-label across 527 sound classes (unbalanced split for pretraining and
downstream fine-tuning; 20k balanced subset for AS-20k).
\textbf{VGGSound}~\cite{chen2020vggsound}: ${\sim}$184k train / 15k test 10\,s
clips spanning 309 audio-visual event classes.

\begin{figure}[t]
  \centering
  \hspace*{\fill}%
  \begin{subfigure}[t]{0.49\linewidth}
    \centering
    \includegraphics[width=\linewidth]{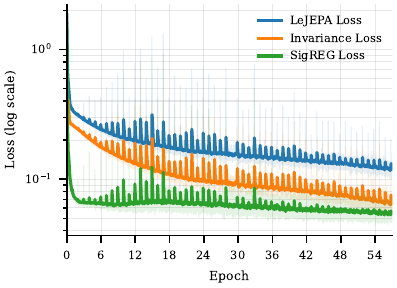}
    \caption{AudioSet-2M (57 ep.)}
    \label{fig:lejepa_loss_audioset}
  \end{subfigure}%
  \hspace*{\fill}%
  \begin{subfigure}[t]{0.49\linewidth}
    \centering
    \includegraphics[width=\linewidth]{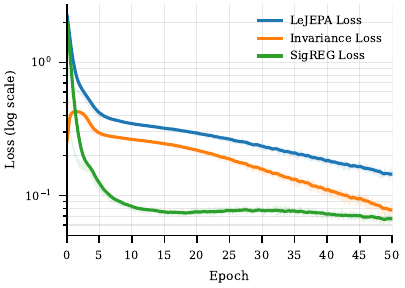}
    \caption{VGGSound (50 ep.)}
    \label{fig:lejepa_loss_vggsound}
  \end{subfigure}%
  \hspace*{\fill}%
  \caption{\textbf{LeJEPA pretraining loss} on {\em (a)}~AudioSet-2M and {\em (b)}~VGGSound,
  decomposed into weighted invariance and SIGReg terms. Both decrease steadily
  without signs of collapse.}
  \label{fig:lejepa_loss}
  \vspace*{-1em}
\end{figure}

\paragraph{Pretraining}
We train \method on 8$\times$ NVIDIA H200 GPUs (DDP, bf16) with
AdamW~\cite{loshchilov2019adamw} at learning rate $5{\times}10^{-4}$, weight
decay 0.05, linear warmup over 15\% of training followed by cosine decay to
$10^{-6}$, gradient clipping 5.0, $G{=}2$ global and $K{=}2$ cross-modal local
views, and $\lambda{=}0.05$. The AudioSet-2M pretraining runs for 57 epochs at
batch size 40/GPU (effective 320); a smaller VGGSound-only run uses 50 epochs at
batch size 50/GPU (effective 400). The pretraining loss decreases monotonically in both regimes
(\cref{fig:lejepa_loss}), and the per-dimension embedding standard deviation
rises from ${\sim}0.8$ to ${\sim}1.01$ on AudioSet (${\sim}0.8$ to ${\sim}1.01$
on VGGSound), confirming that SIGReg converges to the isotropic Gaussian target
at both scales.

\paragraph{Fine-tuning}
We attach a LayerNorm + linear classifier (with an auxiliary attentive head) on
top of \texttt{[CLS]} and unfreeze the full backbone. AdamW with head LR
$2{\times}10^{-4}$ and a low-LR backbone (0.05$\times$ head for the
AS-2M$\to$VGGSound headline, 0.1$\times$ for the controlled and AudioSet runs),
weight decay 0.05, label smoothing 0.1, gradient clipping 1.0, warmup 5\% then
cosine to $10^{-7}$, bf16. We fine-tune VGGSound on 4$\times$ NVIDIA H200 for 13
epochs (6 epochs for the controlled VGGS-only study) and AudioSet on 8$\times$
(AS-2M, ${\sim}29$ epochs) and 2$\times$ (AS-20k, ${\sim}46$ epochs) H200,
reporting best top-1 (resp.\ mAP) along the trajectory with multi-clip
aggregation.

\subsection{Audio-Visual Classification on VGGSound}
\label{sec:exp_vggsound}

\cref{tab:main} compares \method against state-of-the-art audio-visual SSL
methods on VGGSound. After 57 epochs of AudioSet pretraining and 13 epochs of
fine-tuning, \method reaches \textbf{57.1\%} top-1 with the attentive head and
56.6\% with the linear head. To our knowledge, this is the first JEPA-based
result at this level of classification accuracy.

\begin{table}[t]
\caption{Audio-visual classification on VGGSound. \method{} is the only
JEPA-based method. The headline (top) fine-tunes the AS-2M-pretrained backbone
on VGGSound; the middle reports a VGGSound-only controlled study; the bottom
lists published MAE-based baselines.}
\label{tab:main}
\centering\scriptsize
\setlength{\tabcolsep}{6pt}
\begin{tabular}{@{}lclllc@{}}
\toprule
\textbf{Method} & \textbf{Type} & \textbf{Pre-train} &
\textbf{Epochs} & \textbf{Eval} & \textbf{Top-1} \\
\midrule
\multicolumn{6}{@{}l}{\textit{AS-2M $\to$ VGGS fine-tune (ours, headline)}} \\
\method{} (ours) & JEPA & AS-2M & 57+13  & FT (Att.)        & \textbf{57.1} \\
\method{} (ours) & JEPA & AS-2M & 57+13  & FT (Lin.)        & 56.6 \\
\midrule
\multicolumn{6}{@{}l}{\textit{Controlled VGGS-only (ours)}} \\
\method{} (ours) & JEPA & VGGS  & 50+6   & FT               & 49.8 \\
\method{} (ours) & JEPA & VGGS  & 50     & Att.\ (frozen)   & 48.6 \\
\method{} (ours) & JEPA & VGGS  & 50     & Lin.\ (frozen)   & 46.0 \\
\midrule
\multicolumn{6}{@{}l}{\textit{Literature}} \\
MAViL            & MAE  & AS-2M+IN & 80+60  & FT             & 67.1 \\
CAV-MAE          & MAE  & AS-2M    & 25+10  & FT             & 65.4 \\
AV-MAE           & MAE  & VGGS     & 800+50 & FT             & 63.5 \\
CAV-MAE Sync     & MAE  & AS-2M    & 25     & Lin.\ (frozen) & 52.7 \\
\bottomrule
\end{tabular}
\vspace*{-1em}
\end{table}

The end-to-end fine-tuned results of AV-MAE~\cite{georgescu2023avmae},
CAV-MAE~\cite{gong2023cavmae}, and MAViL~\cite{huang2023mavil} (63--67\%) sit
higher, but those methods rely on reconstruction decoders and contrastive
objectives, AV-MAE in particular pretrains for up to $800$ epochs, and MAViL
adds an ImageNet-pretrained visual encoder. The remaining gap is consistent with
these architectural advantages, the absence of a video-specific pretraining
stage, and this being the first JEPA recipe in the audio-visual setting.

\subsection{Audio Classification on AudioSet}
\label{sec:exp_audioset}

\begin{table}[t]
\caption{Audio-visual mAP on the AudioSet eval split. \method{} fine-tunes the
57-epoch AS-2M-pretrained backbone end-to-end on the full AS-2M set (${\sim}29$
epochs) and on the balanced AS-20k subset (${\sim}46$ epochs), and reports
per-modality (audio-only, video-only) eval at inference time. The Epochs column
lists AudioSet-2M pretraining epochs. Baselines report joint A+V mAP.
${}^{\dagger}$linear probe.}
\label{tab:audioset}
\centering\scriptsize
\setlength{\tabcolsep}{4.25pt}
\begin{tabular}{@{}lcllcrr@{}}
\toprule
\textbf{Method} & \textbf{Type} & \textbf{Pre-train} & \textbf{Epochs} & \textbf{Eval} & \textbf{AS-2M} & \textbf{AS-20k} \\
\midrule
\multicolumn{7}{@{}l}{\textit{End-to-end fine-tuning (ours)}} \\
\method{}         & JEPA & AS-2M & 57    &A+V    & \textbf{32.7} & \textbf{29.6} \\
\method{}         & JEPA & AS-2M & 57    &A-only & 26.0 & 23.7 \\
\method{}         & JEPA & AS-2M & 57    &V-only & 12.8 & 10.3 \\
\midrule
\multicolumn{7}{@{}l}{\textit{Baselines (end-to-end fine-tuning)}} \\
AV-MAE             & MAE & AS-2M    & 100 & A+V   & 47.3 & --   \\
CAV-MAE            & MAE & AS-2M    & 25  & A+V   & 51.2 & 42.0 \\
MAViL              & MAE & AS-2M+IN & 80  & A+V   & 53.3 & 44.9 \\
CAV-MAE Sync       & MAE & AS-2M    & --  & A+V   & --   & 30.5${}^{\dagger}$ \\
\bottomrule
\end{tabular}
\vspace*{-1em}
\end{table}

\begin{figure}[t]
  \centering
  \hspace*{\fill}%
  \begin{subfigure}[t]{0.49\linewidth}
    \centering
    \includegraphics[width=\linewidth]{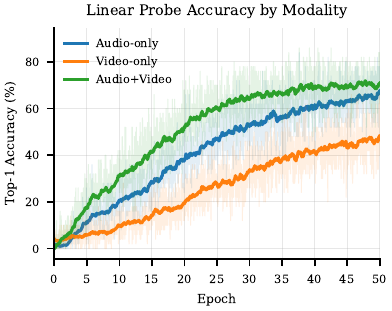}
    \caption{VGGSound pretraining}
    \label{fig:modality_probe_pretrain}
  \end{subfigure}%
  \hspace*{\fill}%
  \begin{subfigure}[t]{0.49\linewidth}
    \centering
    \includegraphics[width=\linewidth]{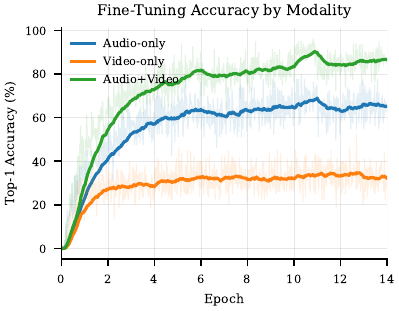}
    \caption{VGGSound fine-tuning}
    \label{fig:modality_probe_ft}
  \end{subfigure}%
  \hspace*{\fill}%
  \caption{\textbf{Top-1 by input modality} during (a)~VGGSound-only pretraining
  (frozen-feature linear probe) and (b)~end-to-end fine-tuning of the AS-2M
  backbone on VGGSound.}
  \label{fig:modality_probe}
  \vspace*{-1em}
\end{figure}

\cref{tab:audioset} reports AudioSet mAP after end-to-end fine-tuning of the
AS-2M-pretrained \method backbone on AS-2M. The model reaches \textbf{32.7 mAP}
on the AudioSet eval split. The per-modality breakdown (26.0 audio-only vs.\
12.8 video-only) shows a clear audio dominance: the JEPA objective produces a
backbone whose decisions are mostly carried by audio, consistent with the
dominant role of audio in many AudioSet classes. The MAE-based baselines reach
$42$--$53$ mAP through dedicated reconstruction decoders and contrastive
objectives (and, for MAViL, an ImageNet-pretrained encoder); closing this gap is
left to future work. The same audio-driven behaviour is visible during VGGSound
training (\cref{fig:modality_probe}): audio-only and audio+video accuracy track
each other closely while video-only lags substantially.

\subsection{Cross-Modal Retrieval and Attention}
\label{sec:retrieval}

\paragraph{Audio$\leftrightarrow$video retrieval}
We probe whether the embedding space is genuinely shared via
audio$\leftrightarrow$video retrieval on the projected \texttt{[CLS]} features.
For each dataset we build a balanced 5-per-class evaluation subset of the
official test split. Each clip is encoded twice through the backbone, once with
the video zeroed (audio-only embedding) and once with the audio zeroed
(video-only embedding), and we rank candidates by cosine similarity in both
A$\to$V (audio query, video gallery) and V$\to$A directions.
\cref{tab:retrieval} reports Recall@$k$ and median rank. On both datasets
retrieval is well above the $1/N$ chance level ($0.06\%$ and $0.05\%$ R@1
respectively), with R@10 reaching ${\sim}36\%$ on both VGGSound and AudioSet,
and the two directions broadly symmetric. These rankings are obtained from the
same backbone with no contrastive training and no paired-retrieval supervision.

\begin{table}[t]
\caption{\textbf{Cross-modal retrieval} on the VGGSound and AudioSet eval splits
(balanced 5-per-class subsets). Recall@$k$ (\%) and median rank in both
directions, on projected \texttt{[CLS]} embeddings.}
\label{tab:retrieval}
\centering\footnotesize
\setlength{\tabcolsep}{8pt}
\begin{tabular}{@{}llcccc@{}}
\toprule
\textbf{Dataset} & \textbf{Dir.} & \textbf{R@1} & \textbf{R@5} & \textbf{R@10} & \textbf{Med.} \\
\midrule
VGGSound            & A$\to$V & 10.61 & 26.34 & 35.40 & 25 \\
($N{=}1545$)        & V$\to$A & 10.16 & 27.38 & 36.89 & 24 \\
\midrule
AudioSet            & A$\to$V & 10.62 & 26.34 & 35.47 & 25 \\
($N{=}2015$)        & V$\to$A & 11.16 & 26.14 & 35.91 & 25 \\
\bottomrule
\end{tabular}
\vspace*{-1em}
\end{table}

\begin{figure}[t]
  \centering
  \setlength{\fboxsep}{1pt}
  \newcommand{\mylabelR}[2]{\tikz[outer sep=0,inner sep=0]{\node[anchor=north west] (a) at (0,0) {#1}; \node[anchor=north west,fill=white,fill opacity=.5,inner sep=2pt,font=\scriptsize] at (a.north west) {#2};}}
  \newcommand{\mylabelL}[2]{\tikz[outer sep=0,inner sep=0]{\node[anchor=north east] (a) at (0,0) {#1}; \node[anchor=north east,fill=white,fill opacity=.5,inner sep=2pt,font=\scriptsize] at (a.north east) {#2};}}
  \begin{minipage}{\columnwidth}
    \centering
    \mylabelL{\includegraphics[height=1.55cm]{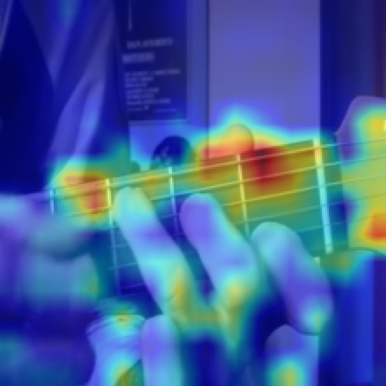}}{V}\hspace{0.5em}%
    \mylabelR{\includegraphics[height=1.55cm,trim=0 0 8pt 0,clip]{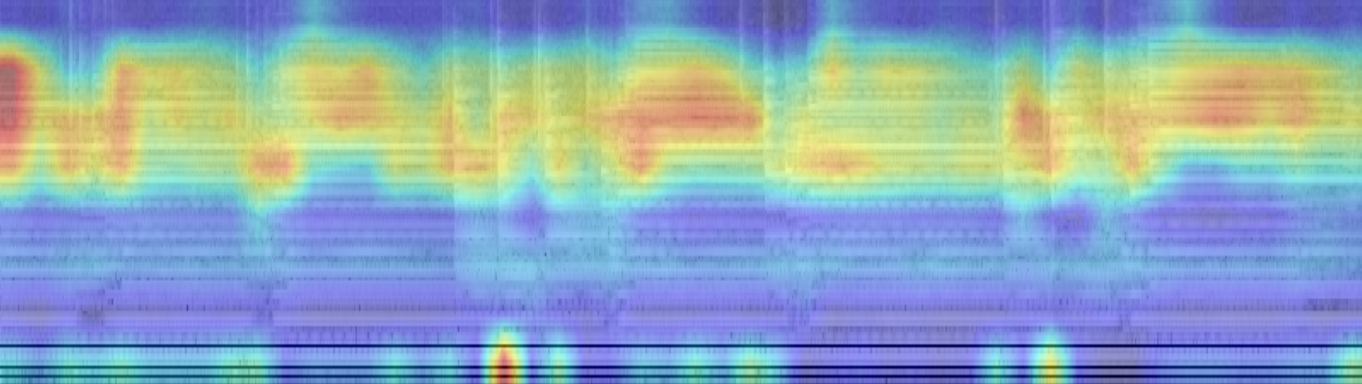}}{A}
  \end{minipage}\\[1pt]
  \begin{minipage}{\columnwidth}
    \centering
    \mylabelL{\includegraphics[height=1.55cm]{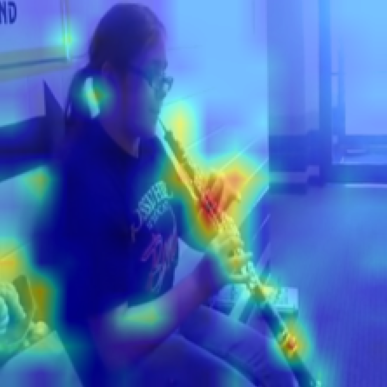}}{V}\hspace{0.5em}%
    \mylabelR{\includegraphics[height=1.55cm,trim=0 0 8pt 0,clip]{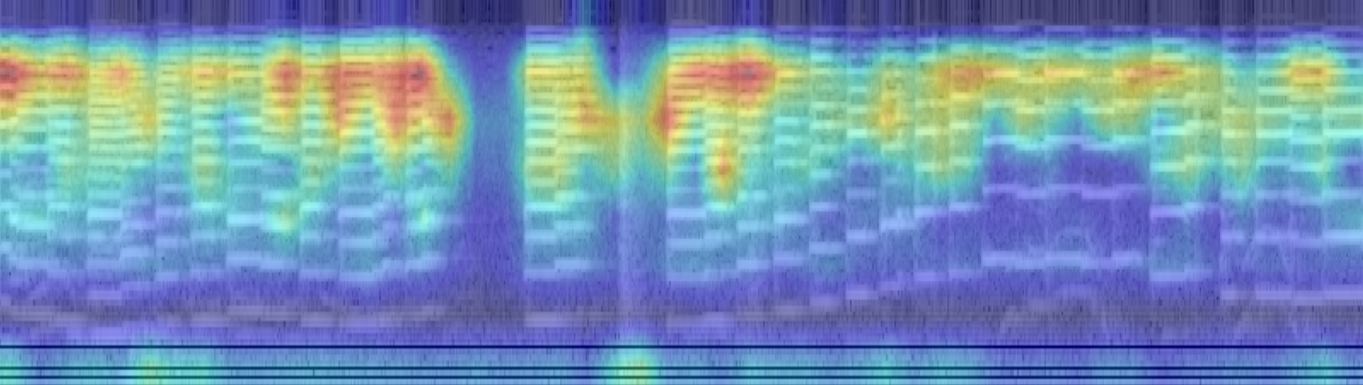}}{A}
  \end{minipage}\\[1pt]
  \begin{minipage}{\columnwidth}
    \centering
    \mylabelL{\includegraphics[height=1.55cm]{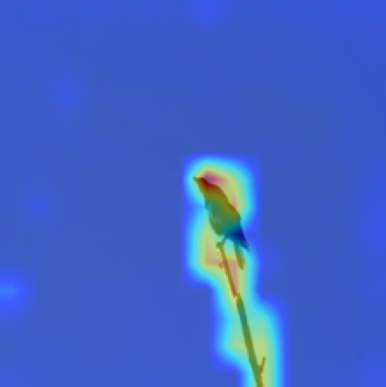}}{V}\hspace{0.5em}%
    \mylabelR{\includegraphics[height=1.55cm]{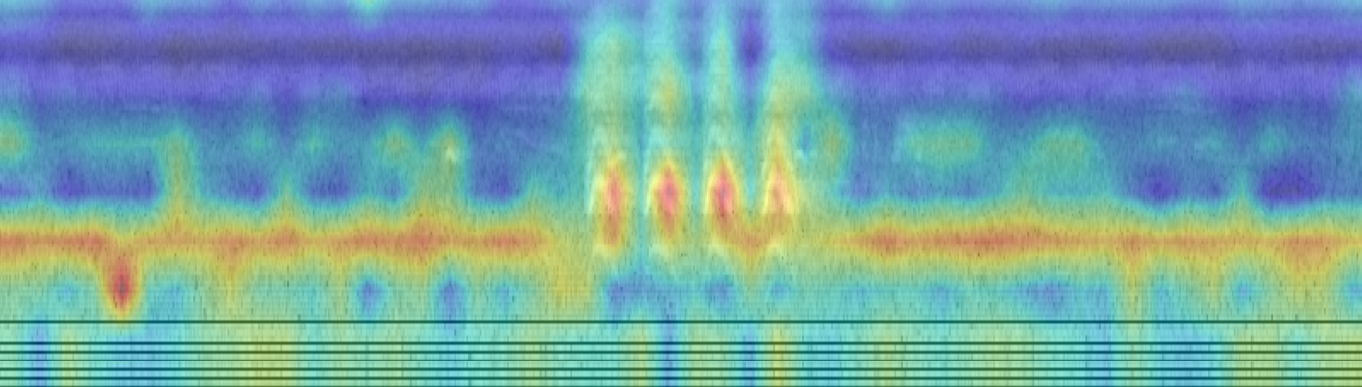}}{A}
  \end{minipage}
  \caption{\textbf{Cross-modal attention} on three VGGSound test clips (guitar,
  flute, bird). Last-layer audio$\to$video attention overlaid on RGB frames (V);
  video$\to$audio attention overlaid on mel spectrograms (A). The model attends
  to the visually salient sound source and to the harmonic / temporal structure
  of the sound, with no localization supervision.}
  \label{fig:attention}
  \vspace*{-1em}
\end{figure}

\paragraph{Cross-modal attention}
\cref{fig:attention} qualitatively visualizes audio$\leftrightarrow$video
attention from the last transformer layer on three VGGSound test clips. We
extract audio-to-video and video-to-audio attention weights, average across
heads, and overlay them on the original RGB frame (audio$\to$video) and on the
mel spectrogram (video$\to$audio). Across all clips the attention concentrates
on the sound-producing region in the video (the guitar and player's hands, the
flute and player's mouth, the body of the flying bird) \emph{and} on the
harmonic / temporal structure in the audio (fundamentals and overtones for the
guitar and flute, the wing-beat envelope for the bird). This emerges from the
JEPA objective alone, with no localization supervision and no contrastive
negatives.\looseness-1

\section{Discussion and Conclusion}
\label{sec:conclusion}

We presented \method, the first extension of LeJEPA to cross-modal audio-visual
self-supervised learning. Replacing spatial masking with \emph{modality
dropout}, \method turns alignment between audio-only and video-only views into
an implicit cross-modal prediction task in latent space, without decoders,
reconstruction targets, stop-gradient, or EMA teachers. The recipe is clean, a
single shared ViT-Base encoder, the LeJEPA loss, and one trade-off scalar
$\lambda$, yet reaches \textbf{57.1\%} top-1 on VGGSound and \textbf{32.7 mAP}
on AudioSet after fine-tuning, and supports zero-shot
audio$\leftrightarrow$video retrieval. This positions theoretically grounded
JEPAs as a viable alternative to masked-autoencoding pipelines for multimodal
representation learning.

\paragraph{Limitations and future work}
Modality-probe and per-modality mAP results show that \method's predictions are
largely carried by audio on both VGGSound and AudioSet, making the backbone most
useful as an \emph{audio} representation with a visual side-channel from
pretraining. The gap with end-to-end fine-tuned MAE-based methods (63--67\% on
VGGSound, 42--53 mAP on AudioSet) likely reflects their reconstruction and
contrastive objectives, the absence of a video-specific pretraining stage, and
(for MAViL) ImageNet initialisation. Next steps include longer AudioSet
pretraining, larger ViTs, initialising the visual stream from a video-only
stage, evaluating against single-modality SSL
backbones~\cite{gong2021ast,huang2022audiomae}, and extending the
modality-dropout recipe to text or optical flow.

\section*{Impact Statement}

This paper presents work whose goal is to advance the field of Machine Learning.
There are many potential societal consequences of our work, none which we feel
must be specifically highlighted here.

\bibliographystyle{icml2026}

\clearpage

\appendix
\section*{Appendices}

\section{LLM/Agent Usage}
\label{app:llm-usage}

The authors made use of LLMs and agents throughout this work. In writing, they
assisted with drafting, editing, and producing plots. They were also used during
model development. LLMs/agents were not used for ideation.

\section{Related Work}
\label{app:related}

\paragraph{Audio-visual SSL}
The dominant approach uses masked autoencoding: AV-MAE~\cite{georgescu2023avmae}
jointly reconstructs masked audio and video; CAV-MAE~\cite{gong2023cavmae}
combines contrastive learning with masked reconstruction;
MAViL~\cite{huang2023mavil} adds self-training; and CAV-MAE
Sync~\cite{araujo2025cavmaesync} introduces fine-grained temporal alignment. All
require decoder networks for pixel/spectrogram reconstruction or contrastive
negatives. \method{} keeps neither.

\paragraph{Audio-only SSL backbones}
On the audio side, AST~\cite{gong2021ast} introduces a transformer backbone for
spectrograms and AudioMAE~\cite{huang2022audiomae} adapts MAE-style masked
spectrogram modelling to audio. \method bypasses per-modality pretraining and
trains a single shared backbone from scratch with a cross-modal JEPA objective.

\paragraph{JEPAs}
I-JEPA~\cite{assran2023ijepa} applies JEPA to images using spatial masking, and
V-JEPA~\cite{bardes2024vjepa} extends it to video.
LeJEPA~\cite{balestriero2025lejepa} provides the theoretical grounding via
SIGReg. All prior JEPA work operates on a single modality; \method is, to our
knowledge, the first to extend JEPA to cross-modal audio-visual learning, with
modality dropout as the partial-view mechanism rather than spatial masking.

\paragraph{Cross-modal alignment}
AudioCLIP~\cite{guzhov2022audioclip} and ImageBind~\cite{girdhar2023imagebind}
learn shared embedding spaces across modalities via large-scale contrastive
pretraining. \method achieves cross-modal alignment as a by-product of the
LeJEPA invariance loss on modality-dropout views, with no contrastive negatives.

\paragraph{Multimodal fusion}
Early fusion concatenates modality tokens before
processing~\cite{nagrani2021mbt}; late fusion uses separate encoders. \method
uses early fusion with modality-type embeddings, enabling the shared transformer
to learn cross-modal interactions from the first layer.

\section{Encoder Architecture}
\label{app:encoder_arch}

\begin{figure*}[p]
  \centering
  \includegraphics[width=\textwidth]{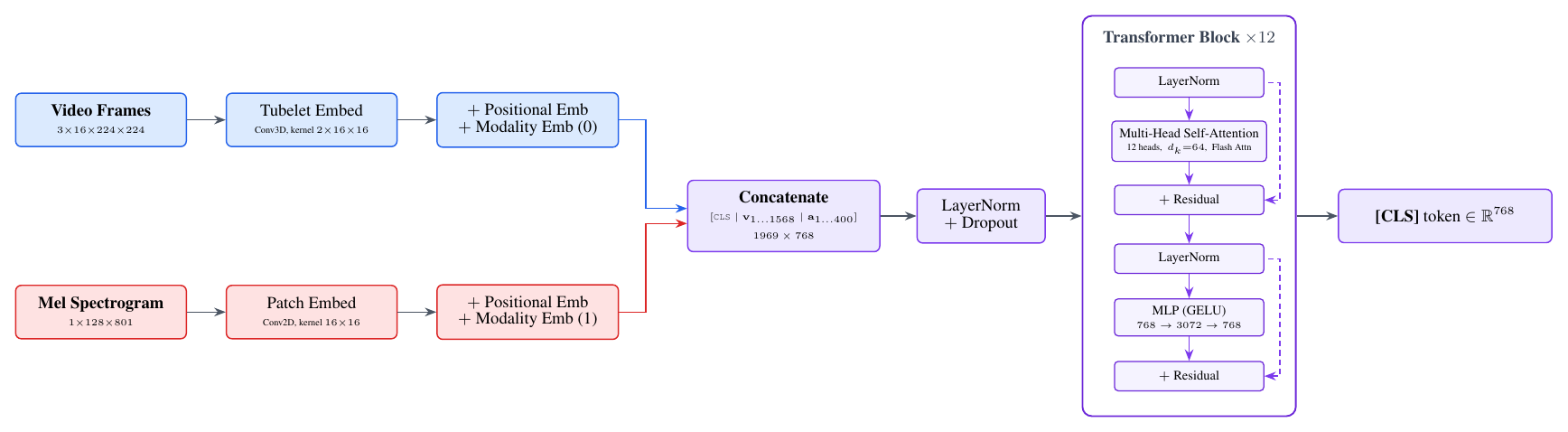}
  \caption{\textbf{\method{} early-fusion encoder.} Video tubelets and
  mel-spectrogram patches are embedded, summed with factorized positional and
  modality-type embeddings, concatenated with a \texttt{[CLS]} token, and
  processed by a $12$-layer ViT-Base. The same encoder is used for every view,
  including the modality-dropout local views, where the absent modality's input
  tensor is zeroed before patch embedding.}
  \label{fig:encoder}
\end{figure*}

\Cref{fig:encoder} shows the full \method{} encoder used for every view (global
and local) during pretraining. Video frames of shape
$3{\times}16{\times}224{\times}224$ are tokenized by a $2{\times}16{\times}16$
Conv3D tubelet into $1568$ video tokens; the mel spectrogram of shape
$1{\times}128{\times}801$ is tokenized by a $16{\times}16$ Conv2D into $400$
audio tokens. Both streams receive factorized positional embeddings (spatial +
temporal for video; frequency + time for audio) and a learnable modality-type
embedding (ID 0 for video, ID 1 for audio). A learnable \texttt{[CLS]} token is
prepended and the resulting $1969$-token sequence is passed through a $12$-layer
ViT-Base ($d{=}768$, $12$ heads, FlashAttention-2). On local views, the tokens
of the dropped modality are zeroed before patch embedding, so the same encoder
handles joint, audio-only, and video-only inputs without architectural changes.
The \texttt{[CLS]} output is taken as the clip embedding and fed to the
projection MLP for the LeJEPA loss (\cref{fig:pipeline}).

\section{Training Configurations}
\label{app:training_configs}

We summarise all training configurations used in the paper in two consolidated
tables. \cref{tab:finetune_config} lists the fine-tuning recipe applied on top
of either backbone for VGGSound classification. \cref{tab:pretrain_config} lists
the pretraining recipes for the AudioSet-2M run (the backbone used in our main
experiment) and the VGGSound-only run (the controlled secondary study).

\section{Additional Embedding-Quality Curves}
\label{app:embed_std}

\begin{figure*}[p]
  \centering
  \begin{subfigure}[t]{0.49\linewidth}
    \centering
    \includegraphics[width=0.8\linewidth]{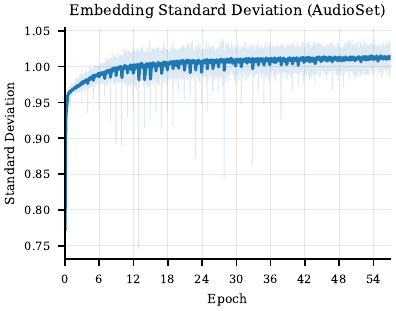}
    \caption{AudioSet-2M (57 ep.)}
  \end{subfigure}\hfill%
  \begin{subfigure}[t]{0.49\linewidth}
    \centering
    \includegraphics[width=0.8\linewidth]{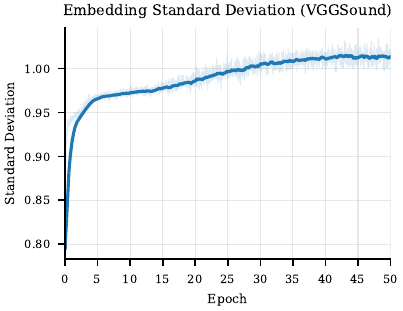}
    \caption{VGGSound (50 ep.)}
  \end{subfigure}
  \caption{\textbf{Embedding standard deviation} over training. SIGReg drives
  the per-dimension std toward 1.0 (the isotropic Gaussian target).}
  \label{fig:embed_std}
\end{figure*}

\cref{fig:embed_std} tracks the per-dimension standard deviation of the
projected embeddings over training for both pretraining runs. In each case the
value rises rapidly toward ${\sim}1.01$ (from ${\sim}0.8$ on VGGSound and
${\sim}0.8$ on AudioSet), confirming that SIGReg drives the embedding
distribution toward the target isotropic Gaussian $\mathcal{N}(\vzero, \mI)$ at
both VGGSound and AudioSet scales.

\section{Additional Pretraining Curves}
\label{app:curves}

\begin{figure*}[p]
\begin{center}
    \begin{subfigure}{0.49\linewidth}
        \centering
        \includegraphics[width=0.8\textwidth]{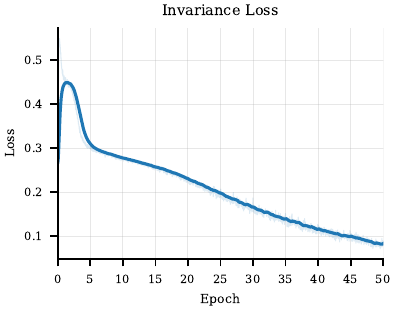}
        \caption{Invariance loss}
        \label{fig:inv_loss}
    \end{subfigure}
    \hfill
    \begin{subfigure}{0.49\linewidth}
        \centering
        \includegraphics[width=0.8\textwidth]{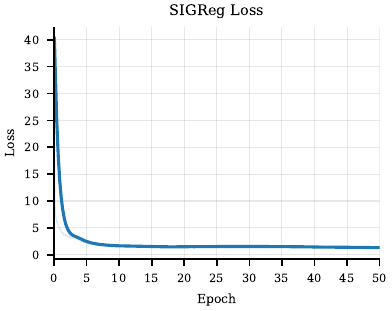}
        \caption{SIGReg loss}
        \label{fig:sigreg_loss}
    \end{subfigure}
    \caption{
        \textbf{Pretraining loss components} (VGGSound-only run, 50 epochs).
    }
    \label{fig:loss_components}
\end{center}
\end{figure*}

\paragraph{Pretraining loss components}
\cref{fig:loss_components} decomposes the LeJEPA objective into its
two unweighted terms.
The invariance term decreases steadily, indicating that view
embeddings converge toward the global center $\bar{\vz}$ and that
cross-modal alignment is successful.
The SIGReg term drops sharply during the first epoch and then
plateaus near zero, confirming that the embedding distribution
matches the target isotropic Gaussian early in training and remains
there for the rest of the run.

\begin{figure*}[p]
    \begin{subfigure}{0.32\linewidth}
        \centering
        \includegraphics[width=0.8\textwidth]{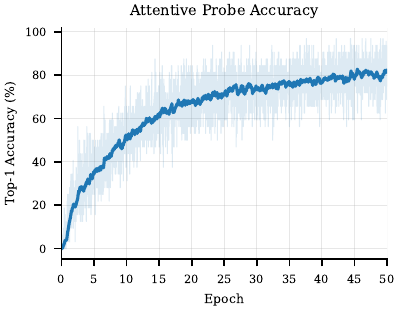}
        \caption{Attentive probe top-1}
        \label{fig:att_probe_acc}
    \end{subfigure}
    \hfill
    \begin{subfigure}{0.32\linewidth}
        \centering
        \includegraphics[width=0.8\textwidth]{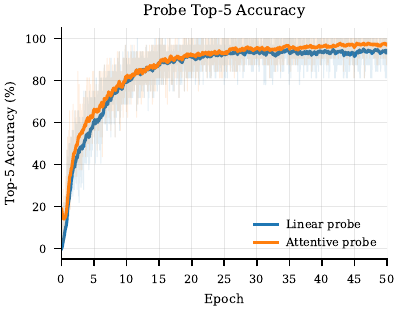}
        \caption{Probe top-5 (linear vs.\ attentive)}
        \label{fig:probe_top5}
    \end{subfigure}
    \hfill
    \begin{subfigure}{0.32\linewidth}
        \centering
        \includegraphics[width=0.8\textwidth]{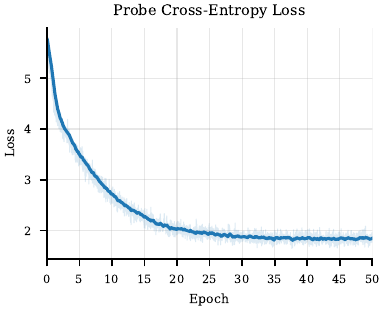}
        \caption{Probe cross-entropy loss}
        \label{fig:probe_loss}
    \end{subfigure}
    \caption{
        \textbf{Online probing curves} on frozen backbone features
        during pretraining.
    }
    \label{fig:probe_curves}
\end{figure*}

\paragraph{Online probe curves}
\cref{fig:probe_curves} reports the online linear and attentive
probes on frozen backbone features during pretraining: attentive
probe top-1 accuracy, top-5 accuracy for both probes, and the
combined cross-entropy loss.
The attentive probe consistently outperforms the linear probe by
${\sim}2$--3 points across both top-1 and top-5.

\section{AudioSet Pretraining Curves}
\label{app:audioset_curves}

\begin{figure*}[p]
\begin{center}
    \begin{subfigure}{0.49\linewidth}
        \centering
        \includegraphics[width=0.8\textwidth]{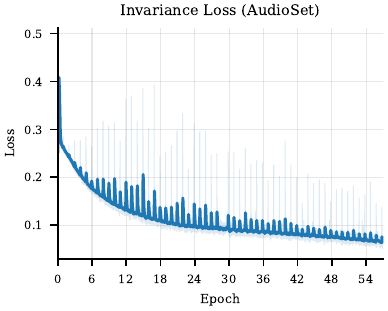}
        \caption{Invariance loss}
        \label{fig:inv_loss_audioset}
    \end{subfigure}
    \hfill
    \begin{subfigure}{0.49\linewidth}
        \centering
        \includegraphics[width=0.8\textwidth]{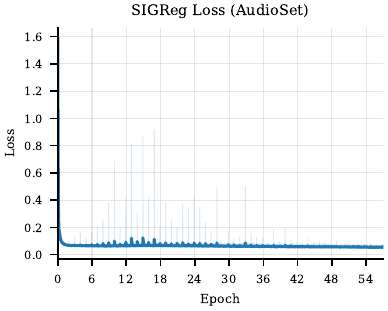}
        \caption{SIGReg loss}
        \label{fig:sigreg_loss_audioset}
    \end{subfigure}
    \caption{
        \textbf{AudioSet pretraining loss components} over 57 epochs of
        AudioSet-2M.
    }
    \label{fig:loss_components_audioset}
\end{center}
\end{figure*}

We complement the AudioSet pretraining loss in the main text with the
per-component invariance and SIGReg curves
(\cref{fig:loss_components_audioset}).
The qualitative behaviour mirrors the VGGSound-only run: the invariance term
decreases steadily and the SIGReg term drops sharply during the first epoch and
then plateaus near zero, confirming that SIGReg behaves as expected at AudioSet
scale.

\section{Fine-Tuning Training Curves}
\label{app:finetune}

\begin{figure*}[p]
\begin{center}
    \begin{subfigure}{0.49\linewidth}
        \centering
        \includegraphics[width=0.8\textwidth]{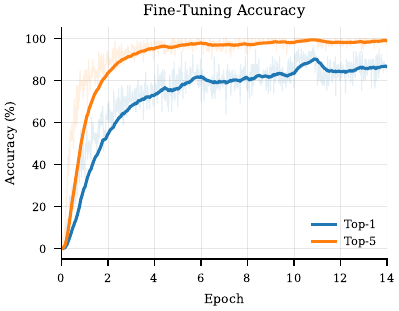}
        \caption{Top-1 / Top-5 accuracy}
        \label{fig:ft_acc}
    \end{subfigure}
    \hfill
    \begin{subfigure}{0.49\linewidth}
        \centering
        \includegraphics[width=0.8\textwidth]{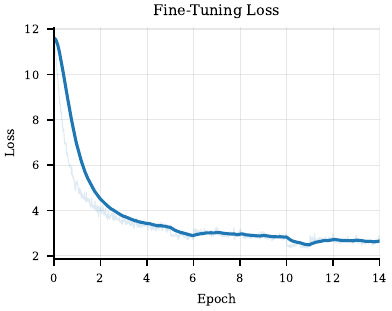}
        \caption{Cross-entropy loss}
        \label{fig:ft_loss}
    \end{subfigure}
    \caption{
        \textbf{Fine-tuning curves} for end-to-end fine-tuning of the
        pretrained \method{} backbone.
    }
    \label{fig:ft_curves}
\end{center}
\end{figure*}

\cref{fig:ft_curves} shows the end-to-end fine-tuning curves
(top-1/top-5 accuracy and cross-entropy loss) when the pretrained
\method{} backbone is fine-tuned with a linear classifier on the
\texttt{[CLS]} token.

\section{Joint Embedding t-SNE by Modality}
\label{app:tsne}

\begin{figure*}[p]
  \centering
  \begin{subfigure}[t]{0.33\linewidth}
    \centering
    \includegraphics[width=0.8\linewidth]{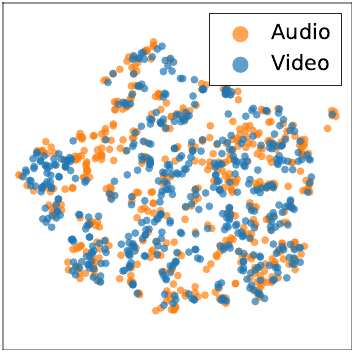}
    \caption{AudioSet}
  \end{subfigure}\hspace{0.03\linewidth}%
  \begin{subfigure}[t]{0.33\linewidth}
    \centering
    \includegraphics[width=0.8\linewidth]{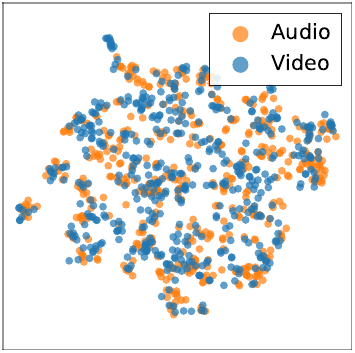}
    \caption{VGGSound}
  \end{subfigure}
  \caption{\textbf{Joint embedding t-SNE by modality.} Projected \texttt{[CLS]}
  embeddings on the 5-per-class retrieval subsets, with each clip contributing
  an audio-only (orange) and a video-only (blue) point. The two modalities are
  mixed rather than separated, indicating a shared cross-modal embedding space.}
  \label{fig:tsne_modality}
\end{figure*}

\cref{fig:tsne_modality} shows two-dimensional t-SNE projections of the
projected \texttt{[CLS]} embeddings on the same balanced 5-per-class evaluation
subsets used for retrieval (AudioSet $N{=}2015$, VGGSound $N{=}1545$ clips).
Each clip contributes two points: one obtained by encoding the clip with the
video tokens zeroed (audio-only, orange) and one with the audio tokens zeroed
(video-only, blue), so an unshared embedding space would manifest as two
well-separated colour clusters. Instead, on both datasets the two modalities are
interleaved across the same regions of the projected space, with no visible
modality-conditioned partition, indicating that \method has learned a genuinely
shared cross-modal embedding space rather than two parallel single-modality
manifolds. This is consistent with the cross-modal retrieval results in
\cref{tab:retrieval}: a shared embedding space is exactly what makes cosine
ranking of audio against video, and vice versa, meaningful in the first place.

\section{Semantic Structure of the Embedding Space}
\label{app:tsne_family}

\begin{figure*}[p]
  \centering
  \includegraphics[width=0.45\textwidth]{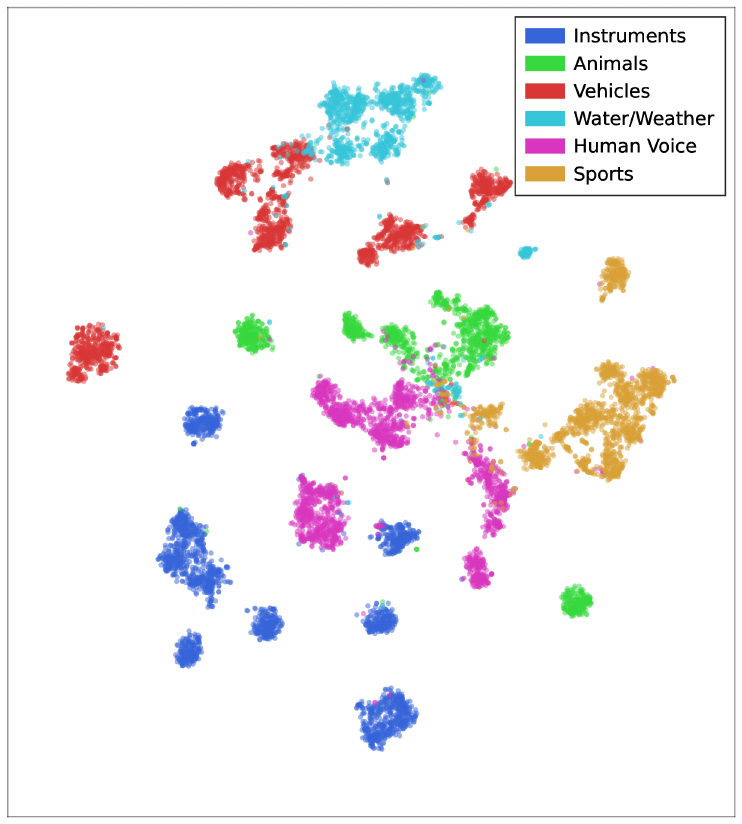}
  \caption{\textbf{Semantic structure of the embedding space.} t-SNE of the
  \texttt{[CLS]} embeddings of the fine-tuned \method{} ViT-B backbone for
  $N{=}11{,}143$ VGGSound training clips drawn from $60$ classes grouped into six semantic
  families (colours). Clips cluster by family with finer per-class sub-structure
  of the same colour; the residual overlap falls mainly between acoustically
  related families (animal calls vs.\ human voice).}
  \label{fig:tsne_family}
\end{figure*}

Where \cref{app:tsne} asks whether the two modalities share a space,
\cref{fig:tsne_family} asks whether that space is organised by semantic content.
We encode $N{=}11{,}143$ VGGSound training clips spanning $60$ classes, grouped into six
coarse semantic families (instruments, animals, vehicles, water/weather, human
voice, and sports), take the \texttt{[CLS]} embedding of each clip from the
fine-tuned \method{} ViT-B backbone (the same backbone behind \cref{tab:main}),
and project the L2-normalised embeddings to two dimensions with t-SNE; points
are coloured by family.
Clips organise into compact, well-separated clusters that respect the family
grouping: each family occupies coherent regions of the projected space, and the
finer per-class structure within a family (for example the individual
instruments) is visible as distinct same-colour sub-clusters.
The residual overlap is concentrated between families that are acoustically and
visually related, most notably animal vocalisations and human voice, which is
consistent with the cross-modal attention behaviour of \cref{fig:attention} and
indicates that the learned space is structured by semantics rather than by
low-level modality cues.

\section{Feature PCA of Video Patch Tokens}
\label{app:pca}

\begin{figure*}[p]
  \centering
  \begin{subfigure}{0.37\linewidth}
    \centering
    \includegraphics[width=\linewidth]{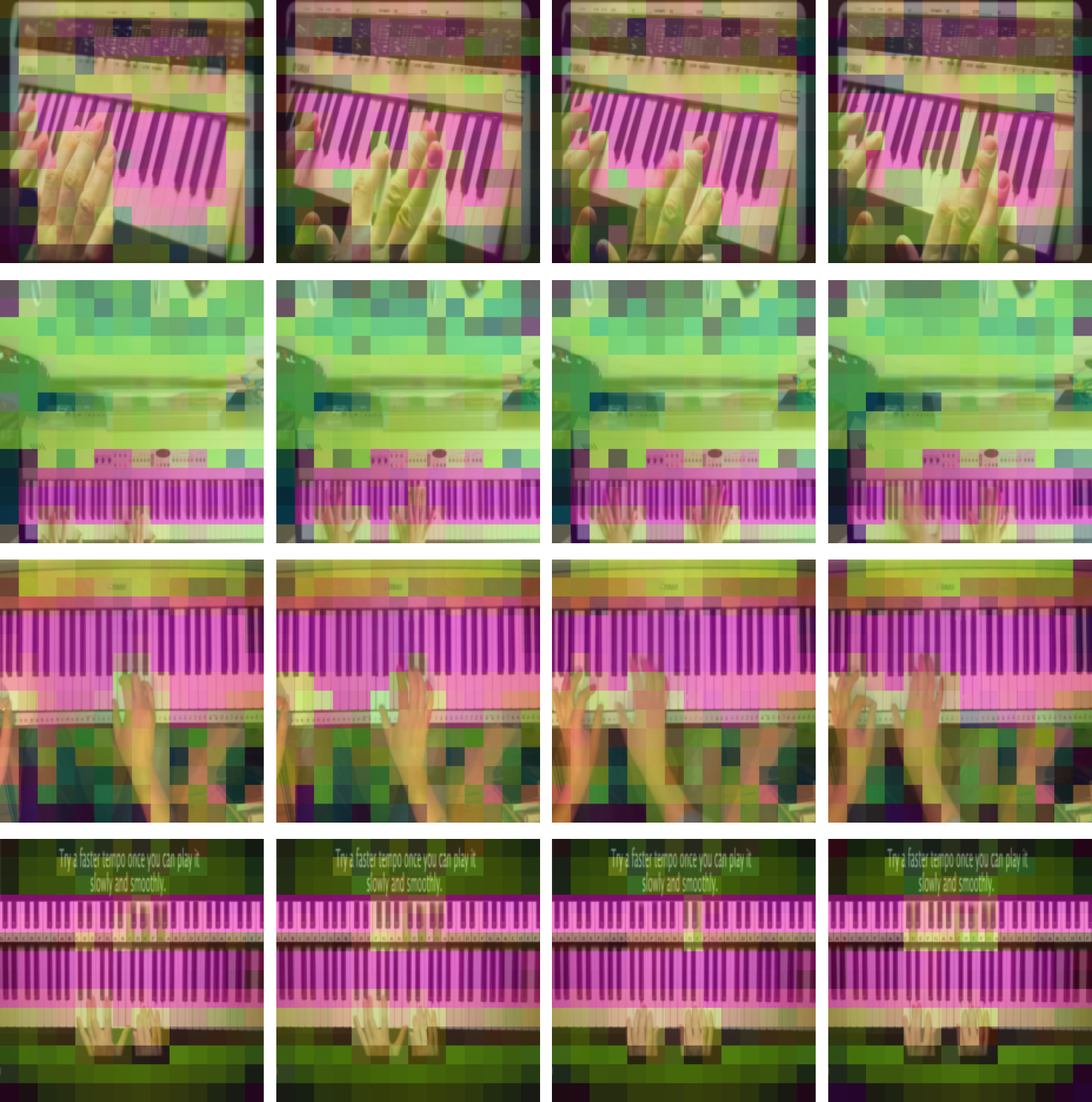}
    \caption{Playing piano}
    \label{fig:pca_piano}
  \end{subfigure}
  \par\smallskip
  \begin{subfigure}{0.37\linewidth}
    \centering
    \includegraphics[width=\linewidth]{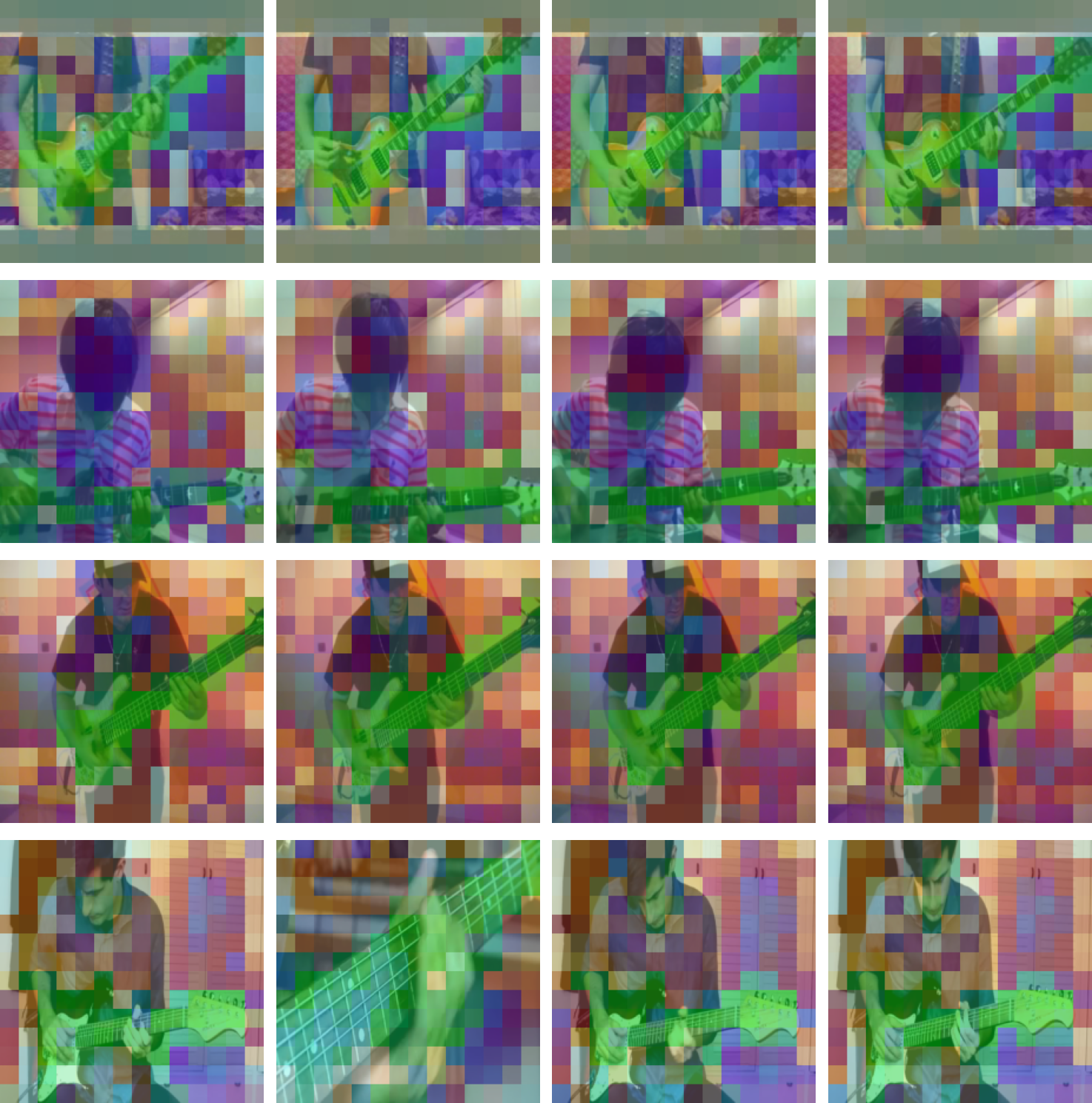}
    \caption{Playing bass guitar}
    \label{fig:pca_bass}
  \end{subfigure}
  \par\smallskip
  \begin{subfigure}{0.37\linewidth}
    \centering
    \includegraphics[width=\linewidth]{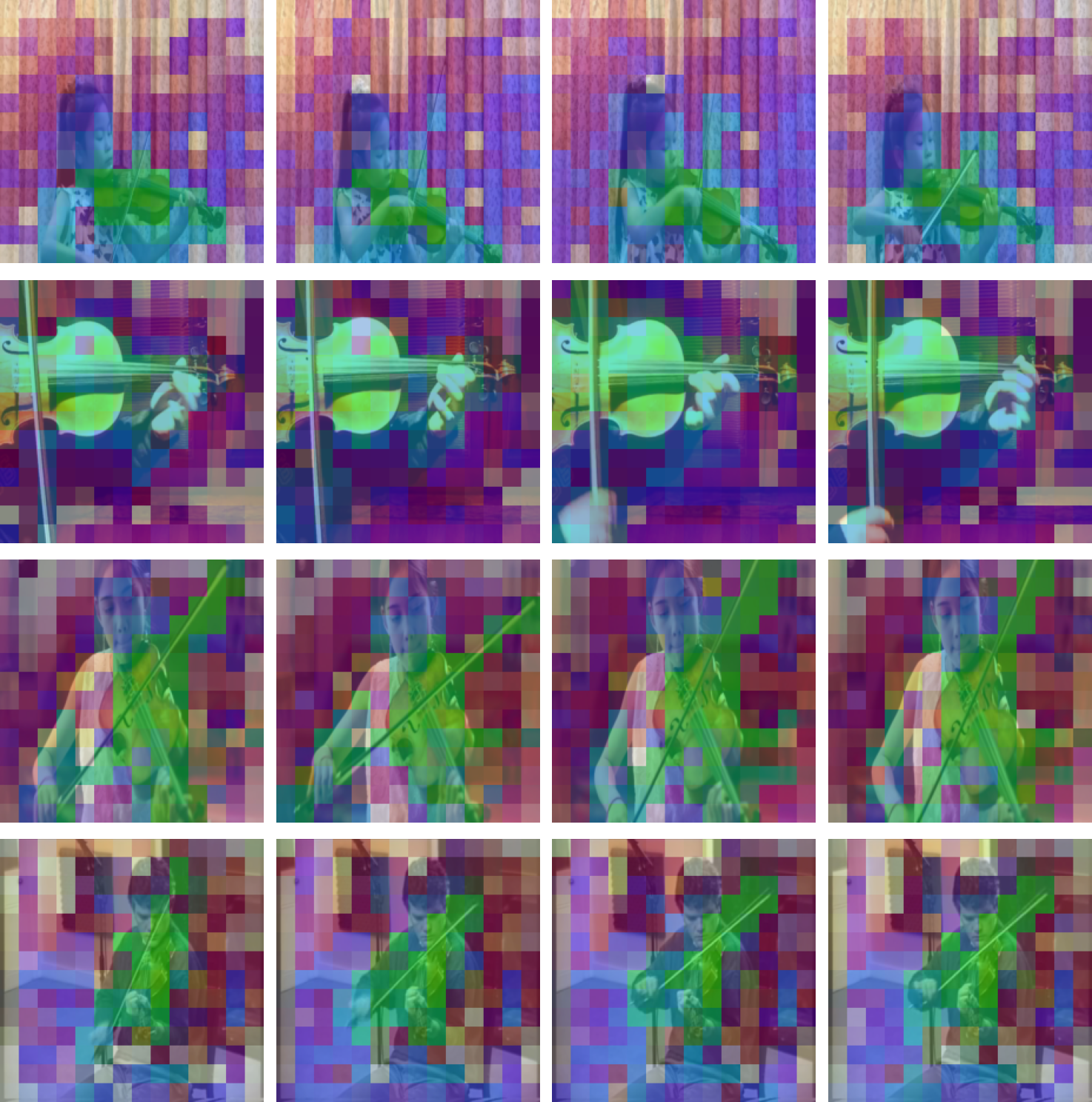}
    \caption{Playing violin}
    \label{fig:pca_violin}
  \end{subfigure}
  \caption{\textbf{Feature PCA of video patch tokens} for three VGGSound
  instrument classes. A single PCA is fit jointly over the last-layer video
  patch tokens of four clips per class (rows); its top three components are
  mapped to RGB and overlaid on four frames per clip (columns). The instrument
  takes a consistent colour across clips and frames, distinct from the player
  and the background.}
  \label{fig:pca}
\end{figure*}

\cref{fig:pca} visualises the video patch-token features of the fine-tuned
\method{} ViT-B encoder (the same backbone behind the VGGSound classification
results in \cref{tab:main}) in the style of DINO/DINOv2 feature PCA.
For each instrument class we fit a \emph{single} PCA jointly over the pooled
last-layer video patch tokens of four VGGSound clips, map the top three
principal components to RGB, and overlay the result on the video frames (rows
are clips, columns are frames; audio tokens are left uncoloured); because the
basis is shared across the four clips, a given colour denotes the same direction
in feature space. The sounding object (the piano keyboard, the body and neck of
the bass guitar, the violin) then takes a consistent colour across clips and
frames, clearly separated from the player and the background, indicating that
the encoder represents the visually salient sound source as a coherent,
instance-corresponding region of feature space, even though it is trained only
with clip-level objectives and never receives pixel- or region-level
supervision, echoing in the video stream the localization behaviour seen in the
cross-modal attention maps of \cref{fig:attention}.

\section{Ablations and LeJEPA Hyperparameter Tuning}
\label{app:ablations}

We complement the main results with one architectural ablation and two
hyperparameter sensitivity studies, all run as VGGSound-only pretraining
with the same ViT-B early-fusion recipe as the controlled run reported in
the main text.
The dual-encoder ablation is run on 8$\times$ NVIDIA H200 to match the
controlled main-text run; the SIGReg-weight $\lambda$ and local-view
$K$ sweeps are run on 1$\times$ NVIDIA H200 (batch size 32 per GPU) under a
tighter compute budget.
All curves are clipped to the shortest variant's final
epoch, ensuring a common training budget.

\begin{figure*}[p]
\centering
\includegraphics[width=0.8\textwidth]{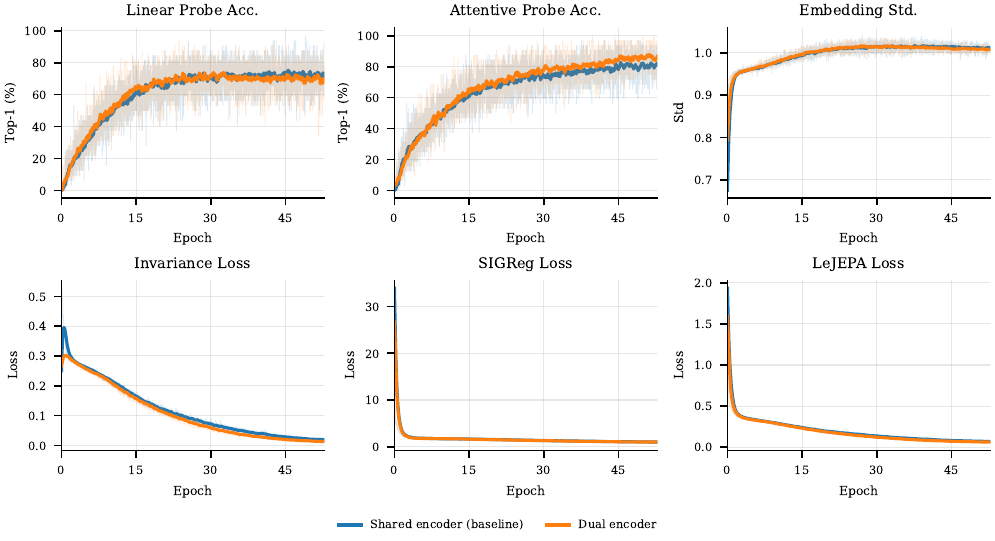}
\caption{
    \textbf{Architectural ablation: shared vs.\ dual encoder} on
    VGGSound pretraining.
    Linear-/attentive-probe top-1 accuracy, embedding standard deviation,
    invariance loss, SIGReg loss, and total LeJEPA loss are plotted
    against pretraining epoch.
    Curves are clipped to the shorter of the two runs ($\sim$53 epochs).
}
\label{fig:abl_dual_encoder}
\end{figure*}

\paragraph{Shared vs.\ dual encoder (ablation).}
\cref{fig:abl_dual_encoder} compares the shared early-fusion encoder used
in \method{} against a dual-encoder variant that
processes audio and video through separate ViT-B encoders and averages the
two \texttt{[CLS]} tokens.
The shared encoder matches the dual-encoder variant on both linear- and
attentive-probe accuracy throughout training; we adopt early fusion for
its simpler, single-encoder design.

\begin{figure*}[p]
\centering
\includegraphics[width=0.8\textwidth]{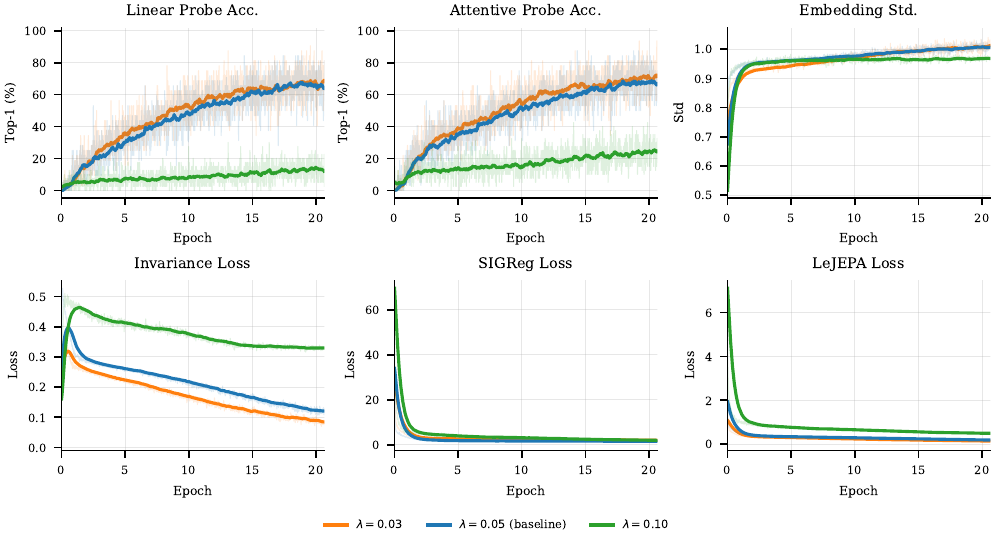}
\caption{
    \textbf{LeJEPA hyperparameter sensitivity: SIGReg weight $\lambda$}
    on VGGSound pretraining, sweeping $\lambda\in\{0.03, 0.05, 0.10\}$
    with $\lambda{=}0.05$ as the baseline used elsewhere in the paper.
    Curves are clipped to the shortest run ($\lambda{=}0.10$,
    $\sim$21 epochs).
}
\label{fig:hp_lambda}
\end{figure*}

\paragraph{SIGReg weight $\lambda$ (LeJEPA hyperparameter).}
\cref{fig:hp_lambda} sweeps the SIGReg weight $\lambda$ in the LeJEPA
loss
$\mathcal{L} = (1-\lambda)\mathcal{L}_{\mathrm{inv}} +
\lambda\mathcal{L}_{\mathrm{SIGReg}}$
across $\{0.03, 0.05, 0.10\}$.
The probe-accuracy curves are tightly clustered, indicating that
\method{} is robust to moderate variation in $\lambda$ on VGGSound.
The SIGReg loss itself naturally rises with $\lambda$ (since the
optimiser tolerates more residual SIGReg in exchange for a stronger
invariance gradient), but the resulting embedding standard deviation
remains close across settings.

\begin{figure*}[p]
\centering
\includegraphics[width=0.8\textwidth]{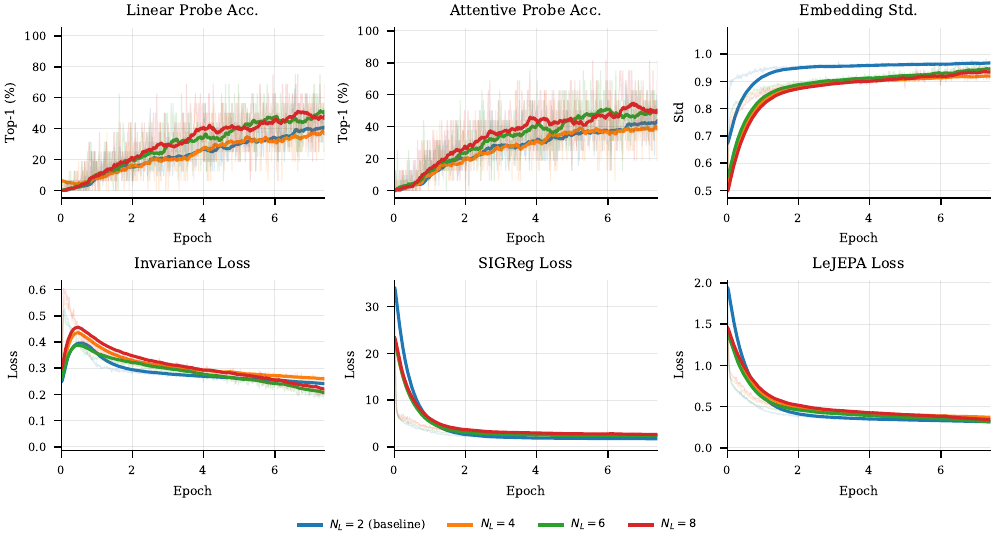}
\caption{
    \textbf{LeJEPA hyperparameter sensitivity: number of local views
    $K$} on VGGSound pretraining, sweeping
    $K\in\{2, 4, 6, 8\}$ with $K{=}2$ as the paper's default.
    Curves are clipped to the shortest run ($K{=}8$, $\sim$7 epochs).
}
\label{fig:hp_local_views}
\end{figure*}

\paragraph{Number of local views $K$ (LeJEPA hyperparameter).}
\cref{fig:hp_local_views} sweeps the number of local (modality-dropout)
views $K$ across $\{2, 4, 6, 8\}$ while keeping the two global views
fixed.
The linear and attentive probes are largely
insensitive to $K$, suggesting that the default of two local views
already provides sufficient cross-modal regularisation; we did not see a
clear benefit from increasing $K$.

\subsection{Loss-component and pipeline ablations}
\label{app:loss_ablations}

We additionally run four hyperparameter ablations of the VGGSound recipe, each
with one change from the 1-GPU ablation baseline (ViT-B, batch size 32,
$\lambda{=}0.05$, modality dropout and video tubelet/frequency-time masking,
attentive probe).
All runs share the same wall-clock budget and reach $\sim$6--7 epochs of
VGGSound pretraining; figures are clipped to the common range.
We deliberately plot only the LeJEPA loss components and the embedding
standard deviation, not probe top-1: with $\lambda{=}0$ the embedding
collapses to a point and with $\lambda{=}1$ views never align, so probe
accuracy is not a meaningful basis for comparison across these runs.
The losses and the embedding distribution remain directly comparable.

\begin{figure*}[p]
\centering
\includegraphics[width=0.8\textwidth]{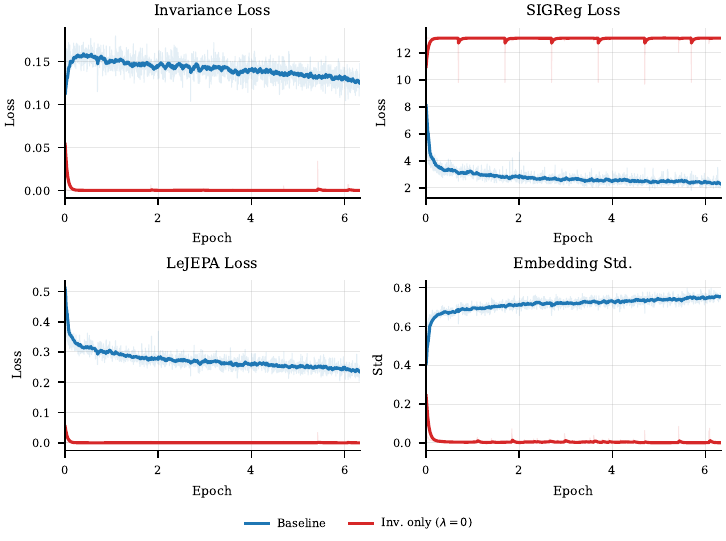}
\caption{
    \textbf{Removing SIGReg ($\lambda{=}0$)} vs.\ the baseline
    ($\lambda{=}0.05$) on VGGSound pretraining.
    Without the SIGReg term, the invariance loss collapses to zero and
    the embedding standard deviation drops several orders of magnitude
    below the baseline, indicating representation collapse.
}
\label{fig:abl_inv_only}
\end{figure*}

\paragraph{Removing the SIGReg term ($\lambda{=}0$).}
\cref{fig:abl_inv_only} sets $\lambda{=}0$, training only on the
invariance loss without the SIGReg regulariser.
The invariance loss drops to ${\sim}10^{-7}$ almost immediately and the
embedding standard deviation collapses from the baseline trajectory
towards ${\sim}10^{-4}$, both consistent with the trivial constant-output
solution.
The SIGReg loss itself, although not optimised, rises far above the
baseline range, confirming that the resulting embedding distribution is
arbitrarily far from the target isotropic Gaussian.
This isolates SIGReg as the component that prevents representation
collapse in the audio-visual setting, mirroring the role it plays in
LeJEPA on images.

\begin{figure*}[p]
\centering
\includegraphics[width=0.8\textwidth]{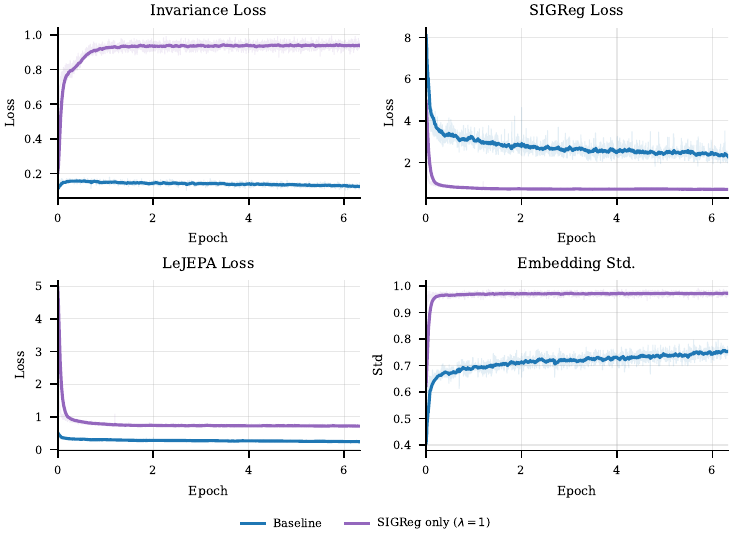}
\caption{
    \textbf{Removing the invariance loss ($\lambda{=}1$)} vs.\ the
    baseline ($\lambda{=}0.05$) on VGGSound pretraining.
    SIGReg alone drives the embedding standard deviation to its target,
    but the invariance loss never decreases, indicating that the model
    never learns to align global and local views without the invariance
    term.
}
\label{fig:abl_sigreg_only}
\end{figure*}

\paragraph{Removing the invariance term ($\lambda{=}1$).}
\cref{fig:abl_sigreg_only} sets $\lambda{=}1$, training only on SIGReg.
The embedding standard deviation reaches the target ${\sim}1$, confirming
that SIGReg alone can shape the distribution, but the invariance loss
stays roughly an order of magnitude above the baseline throughout
training: there is no signal pulling different views of the same clip
together.
This isolates the invariance loss as the component responsible for
cross-modal alignment.

\begin{figure*}[p]
\centering
\includegraphics[width=0.8\textwidth]{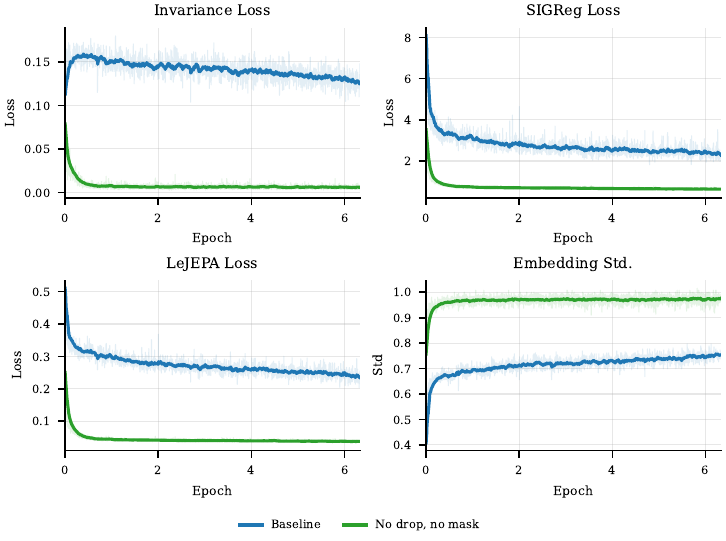}
\caption{
    \textbf{Removing modality dropout and tube/freq-time masking} vs.\
    the baseline on VGGSound pretraining.
    Without partial-view perturbations the invariance loss collapses to
    near zero, indicating that alignment between global and local views
    becomes trivial when both modalities are always present and
    unmasked.
}
\label{fig:abl_no_drop_no_mask}
\end{figure*}

\paragraph{Removing modality dropout and masking.}
\cref{fig:abl_no_drop_no_mask} disables modality dropout, tube masking,
and frequency/time masking ($p_{\mathrm{drop}}{=}0$,
$r_{\mathrm{vid}}{=}0$, $r_{\mathrm{aud}}{=}0$).
Local views then differ from the global views only through random
resized crops and horizontal flips, while always retaining both
modalities.
The invariance loss drops one to two orders of magnitude below the
baseline trajectory because the alignment task is trivialised: there is
no partial-view perturbation forcing the model to predict the missing
modality from the surviving one.
SIGReg behaves similarly to the baseline, so the embedding distribution
is still well shaped, but the LeJEPA loss as a whole loses its
information bottleneck.
This confirms that modality dropout (rather than just spatial cropping)
is what makes the JEPA target non-trivial in the audio-visual setting.

\begin{figure*}[p]
\centering
\includegraphics[width=0.8\textwidth]{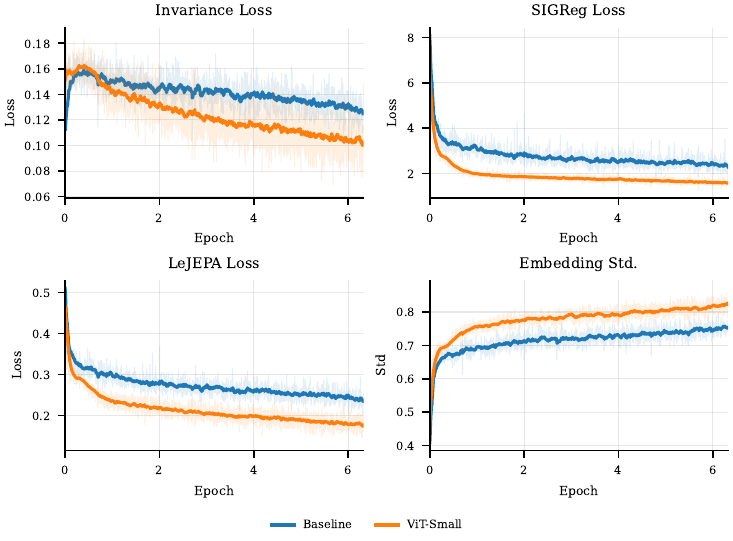}
\caption{
    \textbf{ViT-Small backbone} vs.\ the ViT-Base baseline on VGGSound
    pretraining.
    All three loss components and the embedding standard deviation track
    the baseline curves within a small offset, indicating that the
    LeJEPA recipe transfers to a smaller backbone without retuning.
}
\label{fig:abl_vit_small}
\end{figure*}

\paragraph{Backbone capacity: ViT-Small vs.\ ViT-Base.}
\cref{fig:abl_vit_small} replaces the ViT-B backbone
($d_z{=}768$, ${\sim}86$M parameters) with ViT-S
($d_z{=}384$, ${\sim}22$M parameters), keeping every other
hyperparameter fixed.
ViT-S follows the same overall loss trajectory as the baseline: the
LeJEPA, invariance, and SIGReg losses track the ViT-B curves within a
small constant offset, and the embedding standard deviation converges
toward the same regime.
The smaller backbone reaches a slightly lower invariance loss within the
6--7-epoch budget, plausibly because there is less representational
capacity to spread across, but the qualitative dynamics are unchanged.
This indicates that the LeJEPA recipe is not specific to ViT-B and
transfers to a smaller backbone without retuning $\lambda$ or the view
schedule.

\begin{table*}[p]
\caption{\textbf{Fine-tuning configuration} (VGGSound classification, main
experiment fine-tuning the AudioSet-pretrained backbone for the headline 57.1\%
top-1 result). The controlled VGGSound-only secondary study uses the same recipe
but with 6 epochs.}
\label{tab:finetune_config}
\centering\footnotesize
\setlength{\tabcolsep}{6pt}
\begin{tabular}{@{}ll@{}}
\toprule
\textbf{Setting} & \textbf{Value} \\
\midrule
\multicolumn{2}{@{}l}{\textit{Heads on top of pretrained backbone}} \\
Linear classifier        & LayerNorm + linear on \texttt{[CLS]} (309 classes) \\
Attentive classifier     & 1 query, 12-head cross-attention + linear (309 classes) \\
\midrule
\multicolumn{2}{@{}l}{\textit{Optimization}} \\
Optimizer                & AdamW \\
Head learning rate       & $2{\times}10^{-4}$ \\
Backbone learning rate   & $1{\times}10^{-5}$ ($0.05\times$ head LR) \\
Weight decay             & 0.05 \\
LR schedule              & Linear warmup (5\%) + cosine to $10^{-7}$ \\
Gradient clipping        & 1.0 \\
Label smoothing          & 0.1 \\
Mixed precision          & bf16 \\
Batch size (per GPU)     & 80 \\
Effective batch size     & 320 \\
Epochs                   & 13 \\
\midrule
\multicolumn{2}{@{}l}{\textit{Data and views}} \\
SSL augmentations        & none (single global view, no modality dropout) \\
Train-time eval          & best top-1 along trajectory \\
Test-time aggregation    & 6 clips, averaged logits \\
\midrule
\multicolumn{2}{@{}l}{\textit{Compute}} \\
GPUs                     & 4$\times$ NVIDIA H200 \\
Parallelism              & DDP \\
Wall-clock time          & ${\sim}29$\,h \\
\bottomrule
\end{tabular}
\end{table*}

\begin{table*}[p]
\caption{\textbf{Pretraining configurations.} Both runs share the same backbone
(ViT-Base early-fusion), optimizer, and LeJEPA recipe, and differ only in
dataset, batch size, and number of epochs. The online linear and attentive
classification probes are attached only to the VGGSound-only run.}
\label{tab:pretrain_config}
\centering\footnotesize
\setlength{\tabcolsep}{10pt}
\begin{tabular}{@{}lll@{}}
\toprule
\textbf{Setting} & \textbf{AudioSet-2M (main)} & \textbf{VGGSound (controlled)} \\
\midrule
\multicolumn{3}{@{}l}{\textit{Data}} \\
Dataset                  & AudioSet-2M (unbalanced)        & VGGSound (train) \\
Training samples         & ${\sim}$2M                       & ${\sim}$184k \\
Clip length              & 8\,s (from 10\,s clip, random offset) & 8\,s (from 10\,s clip, random offset) \\
\midrule
\multicolumn{3}{@{}l}{\textit{Architecture}} \\
Encoder                  & ViT-Base (12L, 768d, 12h)        & ViT-Base (12L, 768d, 12h) \\
Audio embed              & Conv2D, kernel/stride $16{\times}16$ & Conv2D, kernel/stride $16{\times}16$ \\
Video embed              & Conv3D, kernel/stride $2{\times}16{\times}16$ & Conv3D, kernel/stride $2{\times}16{\times}16$ \\
Sequence length          & 1969 (1 CLS + 1568 V + 400 A)    & 1969 (1 CLS + 1568 V + 400 A) \\
Projector                & 3-layer MLP, $768{\to}2048{\to}2048{\to}128$ & 3-layer MLP, $768{\to}2048{\to}2048{\to}128$ \\
\midrule
\multicolumn{3}{@{}l}{\textit{LeJEPA recipe}} \\
Global views $G$         & 2                                & 2 \\
Local views $K$          & 2 (cross-modal: 1 audio-only, 1 video-only) & 2 (cross-modal: 1 audio-only, 1 video-only) \\
Modality dropout         & yes                              & yes \\
$\lambda$ (SIGReg weight) & 0.05                            & 0.05 \\
SIGReg directions        & resampled per step               & resampled per step \\
\midrule
\multicolumn{3}{@{}l}{\textit{Optimization}} \\
Optimizer                & AdamW                            & AdamW \\
Learning rate            & $5{\times}10^{-4}$               & $5{\times}10^{-4}$ \\
Weight decay             & 0.05                             & 0.05 \\
LR schedule              & Linear warmup (15\%) + cosine to $10^{-6}$ & Linear warmup (15\%) + cosine to $10^{-6}$ \\
Gradient clipping        & 5.0                              & 5.0 \\
Mixed precision          & bf16                             & bf16 \\
Batch size (per GPU)     & 40                               & 50 \\
Effective batch size     & 320                              & 400 \\
Epochs                   & 57                               & 50 \\
Optimizer steps          & ${\sim}341$k                     & ${\sim}23$k \\
\midrule
\multicolumn{3}{@{}l}{\textit{Online probes (frozen backbone, gradients detached)}} \\
Linear probe             & --                               & LayerNorm + linear on \texttt{[CLS]} \\
Attentive probe          & --                               & 1 query, 12-head cross-attention \\
Probe LR                 & --                               & $10^{-3}$ \\
Probe weight decay       & --                               & 0 \\
Probe label smoothing    & --                               & 0.1 \\
\midrule
\multicolumn{3}{@{}l}{\textit{Compute}} \\
GPUs                     & 8$\times$ NVIDIA H200            & 8$\times$ NVIDIA H200 \\
Parallelism              & DDP                              & DDP \\
Wall-clock time          & ${\sim}192$\,h                   & ${\sim}10$\,h \\
\bottomrule
\end{tabular}
\end{table*}

\clearpage

\end{document}